\documentclass[11pt]{article}

\usepackage[dvips]{graphicx}

\usepackage{graphics}

\newcommand{\lrD}{{D^{\hspace{-0.8em}
      \raisebox{0.8ex}{$\scriptstyle\leftrightarrow$}}}{}}
\newcommand{\lD}{{D^{\hspace{-0.8em}
      \raisebox{0.8ex}{$\scriptstyle\leftarrow$}}}{}}
\newcommand{\rD}{{D^{\hspace{-0.8em}
      \raisebox{0.8ex}{$\scriptstyle\rightarrow$}}}{}}

\newcommand{\lpartial}{{\partial^{\hspace{-0.65em}
      \raisebox{0.8ex}{$\scriptstyle\leftarrow$}}}{}}
\newcommand{\rpartial}{{\partial^{\hspace{-0.65em}
      \raisebox{0.8ex}{$\scriptstyle\rightarrow$}}}{}}

\newcommand{\stackdown}[2]{\mathrel{\mathop{#1}\limits_{#2}}}

\newcommand{\fslash}[1]{{}\hspace{-1.0ex} \not{\hspace{-1.2ex}#1}}

\newcommand{\tvec}[1]{\mbox{\boldmath{$#1$}}}
\newcommand{\svec}[1]{\mbox{\boldmath{$\scriptstyle #1$}}}

\newcommand{\half}{{\textstyle\frac{1}{2}}}

\newcommand{\mm}{m_{\hspace{-0.5pt}A\hspace{0.6pt}}}

\begin{document}

\begin{flushright}
DESY 05-061 \\
hep-ph/0504175 \\
\end{flushright}

\begin{center}
\vskip 4.0\baselineskip
{\LARGE \bf Spin densities in the transverse plane and \\[0.3em]
   generalized transversity distributions }
\vskip 5.0\baselineskip
M.~Diehl \\[0.5\baselineskip]
\textit{Deutsches Elektronen-Synchroton DESY, 22603 Hamburg, Germany}
\\[1.5\baselineskip] 
Ph.~H{\"a}gler \\[0.5\baselineskip]
\textit{Department of Physics and Astronomy, Vrije Universiteit, De
  Boelelaan 1081, \\ 1081 HV Amsterdam, The Netherlands}
\vskip 6.0\baselineskip
\textbf{Abstract}\\[0.5\baselineskip]
\parbox{0.9\textwidth}{We show how generalized quark distributions in
  the nucleon describe the density of polarized quarks in the impact
  parameter plane, both for longitudinal and transverse polarization
  of the quark and the nucleon.  This density representation entails
  positivity bounds including chiral-odd distributions, which tighten
  the known bounds in the chiral-even sector.  Using the quark
  equations of motion, we derive relations between the moments of
  chiral-odd generalized parton distributions of twist two and twist
  three.  We exhibit the analogy between polarized quark distributions
  in impact parameter space and transverse momentum dependent
  distribution functions.}
\end{center}
\vskip 3.0\baselineskip


\section{Introduction}

The distribution of transverse quark spin in the proton remains one of
the most intriguing and least known aspects of nucleon structure.  It
has been the subject of numerous theoretical studies, and there is a
vigorous experimental program aiming to measure the transversity
distribution $h_1(x)$ in present or planned experiments.  Recent
overviews and references can for instance be found in
\cite{Vogelsang:2003eb}.

A wealth of information on the nucleon structure is encoded in
generalized parton distributions (GPDs), see e.g.\ the reviews
\cite{Goeke:2001tz,Diehl:2003ny,Belitsky:2005qn}.  They admit a
particularly intuitive physical interpretation at zero skewness $\xi$,
where after a Fourier transform they describe how partons with given
longitudinal momentum are spatially distributed in the transverse
plane \cite{Burkardt:2000za}.  A remarkable spin effect in this
representation is that transverse nucleon polarization induces a
sideways shift in the quark density, whose size is related to the
anomalous magnetic moment of the nucleon and thus quite
substantial~\cite{Burkardt:2002hr}.

A relatively small number of studies have so far been devoted to
generalized transversity distributions, which were introduced in
\cite{Collins:1997fb,Hoodbhoy:1998vm,Diehl:2001pm}.  Since the
operator measuring transversity is chiral-odd, it is notoriously
difficult to find processes where transversity distributions can be
accessed experimentally.  For generalized transversity distributions
it is indeed not clear if this can be achieved in practice, and at
present there is only one type of process known where this may be
possible in principle \cite{Ivanov:2002jj}.  There is however the
prospect of gaining information from lattice QCD, which provides a
tool to calculate the Mellin moments of generalized parton
distributions.  Several studies have been performed for chiral-even
distributions \cite{Gockeler:2003jf,Hagler:2003jd}, and first results
for chiral-odd ones have been presented in \cite{Gockeler:2005aw}.
The purpose of this paper is to take a closer look at the physical
interpretation and properties of these quantities.

In Sect.~\ref{sec:transv} we extend the analysis of
\cite{Burkardt:2002hr} to generalized transversity distributions and
investigate the distribution of transverse quark polarization in the
impact parameter plane.  The result closely resembles the expressions
for the distribution of polarized quarks as a function of their
transverse momentum.  In Sect.~\ref{sec:positive} we derive positivity
bounds which involve chiral-even and chiral-odd distributions, both in
impact parameter and in momentum representation.  We also give bounds
that are valid for Mellin moments.  Sect.~\ref{sec:eom} is devoted to
relations between distributions of twist two and three resulting from
the QCD equations of motion.  We summarize our findings in
Sect.~\ref{sec:sum}.


\section{Polarized parton distributions in the transverse plane}
\label{sec:transv}

To begin with, let us recall the definitions for generalized quark
distributions in the proton.  Following the conventions of
\cite{Ji:1998pc,Diehl:2001pm,Diehl:2003ny} the distributions of twist
two read
\begin{eqnarray}
  \label{gpd-def}
F(x,\xi,t) &=& \int \frac{dz^-}{4\pi}\, e^{i x P^+ z^-}\,
    \langle p'|\, \bar{q}(-\half z) \gamma^+ q(\half z) |\,
    p\rangle \,\Big|_{z^+ =0,\, \svec{z}=0} 
\nonumber \\[0.2em]
 &=& \frac{1}{2P^+} \Bigg[
  H(x,\xi,t)\, \bar{u} \gamma^+ u +
  E(x,\xi,t)\, \bar{u} \frac{i \sigma^{+\alpha} \Delta_\alpha}{2m} u
  \, \Bigg] ,
\nonumber \\[0.2em]
\tilde{F}(x,\xi,t) &=& \int \frac{dz^-}{4\pi}\, e^{i x P^+ z^-}\,
    \langle p'|\, \bar{q}(-\half z) \gamma^+ \gamma_5\, q(\half z) |\,
    p\rangle \,\Big|_{z^+ =0,\, \svec{z}=0} 
\nonumber \\[0.2em]
 &=& \frac{1}{2P^+} \Bigg[
  \tilde{H}(x,\xi,t)\, \bar{u} \gamma^+ \gamma_5 u +
  \tilde{E}(x,\xi,t)\, \bar{u} \frac{\gamma_5 \Delta^+}{2m} u
  \, \Bigg] ,
\nonumber \\[0.2em]
F^{j}_T(x,\xi,t) &=& -i \int \frac{dz^-}{4\pi}\, e^{i x P^+ z^-}\,
    \langle p'|\, \bar{q}(-\half z)\, \sigma^{+j} \gamma_5\,
    q(\half z) |\, p\rangle \,\Big|_{z^+ =0,\, \svec{z}=0} 
\nonumber \\[0.2em]
 &=& -\frac{i}{2P^+} \Bigg[
  H_T(x,\xi,t)\, \bar{u} \sigma^{+j} \gamma_5\, u 
  + \tilde{H}_T(x,\xi,t)\, \bar{u} \frac{\epsilon^{+j\alpha\beta}
    \Delta_\alpha P_\beta}{m^2} u
\nonumber \\
 && \hspace{2em}
{}+ E_T(x,\xi,t)\, \bar{u} \frac{\epsilon^{+j\alpha\beta}
    \Delta_\alpha \gamma_\beta}{2m} u
  + \tilde{E}_T(x,\xi,t)\, \bar{u} \frac{\epsilon^{+j\alpha\beta}
    P_\alpha \gamma_\beta}{m} u 
  \, \Bigg] .
\end{eqnarray}
Corresponding to the quark-antiquark operator in their definition, the
distributions parameterizing $F$ and $\tilde{F}$ are referred to as
chiral-even, and those parameterizing $F_T^j$ as chiral-odd.  The
latter are also called quark helicity flip or generalized transversity
distributions.  We use light-cone coordinates $v^{\pm} = (v^0 \pm v^3)
/\sqrt{2}$ for any four-vector $v$ and write its transverse part as
$\tvec{v} = (v^1, v^2)$.  Scalar products of boldface vectors are
defined such that $\tvec{v}^2 \ge 0$, and Roman indices $i$, $j$, $k$
are understood to be restricted to the two transverse directions.  Our
sign convention for the totally antisymmetric tensor is
$\epsilon_{0123} = 1$.  We use kinematical variables $P = \half
(p+p')$, $\Delta = p'-p$, $t = \Delta^2$, $\xi = (p-p')^+ /(p+p')^+$
and denote the proton mass by $m$.  For better legibility we have not
explicitly labeled the polarization of the proton states $\langle p'|$
and $|p\rangle$ and have omitted the momentum and polarization labels
of the proton spinors $\bar{u}$ and $u$.  The definitions
(\ref{gpd-def}) hold in the light-cone gauge $A^+ =0$, otherwise a
Wilson line appears between the quark field and its conjugate.

We will find that in all expressions of this paper the distribution
$E_T$ appears in the combination $E_T + 2\tilde{H}_T$, so that one may
regard $E_T + 2\tilde{H}_T$ as a more fundamental quantity than $E_T$.
Using the Gordon identity one can make this combination appear already
in the decomposition of the matrix element $F_T^j(x,\xi,t)$, rewriting
\begin{eqnarray}
\lefteqn{
H_T\, \bar{u} \sigma^{+j} \gamma_5\, u 
  + \tilde{H}_T\, \bar{u} \frac{\epsilon^{+j\alpha\beta}
    \Delta_\alpha P_\beta}{m^2} u
  + E_T\, \bar{u} \frac{\epsilon^{+j\alpha\beta}
    \Delta_\alpha \gamma_\beta}{2m} u
}
\\
&=& H_T\, \bar{u} \sigma^{+j} \gamma_5\, u 
  - \tilde{H}_T\, \bar{u} \frac{\epsilon^{+j\alpha\beta}\,
    \Delta_\alpha\, i\sigma_{\beta\delta\,} \Delta^\delta}{2m^2} u
  + (E_T + 2\tilde{H}_T)\, \bar{u} \frac{\epsilon^{+j\alpha\beta}
    \Delta_\alpha \gamma_\beta}{2m} u
\nonumber \\
&=& \Big( H_T - \frac{t}{2m^2}\tilde{H}_T \Big)\,
    \bar{u} \sigma^{+j} \gamma_5\, u 
  + \tilde{H}_T\, \bar{u} \frac{
      \Delta^j \sigma^{+\alpha}\gamma_5 \Delta_\alpha
    - \Delta^+ \sigma^{j\alpha}\gamma_5 \Delta_\alpha}{2m^2} u
  + (E_T + 2\tilde{H}_T)\, \bar{u} \frac{\epsilon^{+j\alpha\beta}
    \Delta_\alpha \gamma_\beta}{2m} u \, .
\nonumber
\end{eqnarray}

In this and the next section we restrict ourselves to skewness
$\xi=0$, where generalized parton distributions have a probability
interpretation when transformed to impact parameter space
\cite{Burkardt:2000za}.  To make this explicit we form wave packets
\begin{equation}
  \label{impact-states}
|p^+, \tvec{b}\rangle = 
  \int \frac{d^2 \tvec{p}}{(2\pi)^2}\, e^{-i\svec{b} \svec{p}}\, 
  | p\rangle
\end{equation}
from the states states $|p\rangle$ with definite four-momentum, where
it is understood that the integration over $\tvec{p}$ is done at fixed
$p^+$ with $p^- = (m^2 + \tvec{p}^2)/(2p^+)$.  The state $|p^+,
\tvec{b}\rangle$ has definite plus-momentum $p^+$ and definite impact
parameter $\tvec{b}$, i.e., it is localized at position $\tvec{b}$ in
the $x$-$y$ plane.  Further analysis shows that $\tvec{b}$ is the
``center of momentum'' of the partons in the proton
\cite{Soper:1976jc}, given in terms of their plus-momenta and
transverse positions as $\tvec{b} = \sum_i p^+_i
\tvec{b}_i^{\phantom{+}} /\sum_i p^+_i$.  A two-dimensional Fourier
transform gives
\begin{eqnarray}
  \label{impact-mat-elements}
F(x,\tvec{b}) &=&
  \int \frac{d^2 \tvec{\Delta}}{(2\pi)^2}\, 
     e^{- i\svec{b} \svec{\Delta}}\, F(x,0,-\tvec{\Delta}^2)
\nonumber \\
 &=& \mathcal{N}^{-1}\, \int \frac{dz^-}{4\pi}\, e^{i x p^+ z^-}\,
  \Big\langle p^+,\tvec{0} \,\Big|\, \bar{q}(z_2) \gamma^+
  q(z_1) \Big|\, p^+,\tvec{0} \Big\rangle
\end{eqnarray}
with $z_1^+ =z_2^+ =0$, $\tvec{z}_1 =\tvec{z}_2 = \tvec{b}$, and
$z_1^- =-z_2^- =\half z^-$.  Here we have used translation invariance
to shift the quark-antiquark operator to transverse position
$\tvec{b}$ and the impact parameter of the proton to the origin.  The
normalization factor $\mathcal{N} = (2\pi)^{-2} \int d^2\tvec{p}$ is
singular like a delta-function, which can be avoided if instead of
(\ref{impact-states}) one takes wave packets smeared out in impact
parameter space \cite{Burkardt:2000za,Diehl:2002he}.  In analogy to
(\ref{impact-mat-elements}) we define matrix elements
$\tilde{F}(x,\tvec{b})$ and $F_T^j(x,\tvec{b})$.  The impact parameter
distribution
\begin{equation}
  \label{impact-gpds}
H(x,\tvec{b}^2) = \int \frac{d^2 \tvec{\Delta}}{(2\pi)^2}\, 
     e^{- i\svec{b} \svec{\Delta}}\, H(x,0,-\tvec{\Delta}^2)
\end{equation}
and its analogs $E(x,\tvec{b}^2)$, $\tilde{H}(x,\tvec{b}^2)$,
$H_T(x,\tvec{b}^2)$, etc.\ depend on $\tvec{b}$ only via its square
thanks to rotation invariance.  We see in (\ref{impact-mat-elements})
that the Fourier transformation has made the matrix element diagonal
in the plus-momentum and the impact parameter of the proton states.
If we also take the same polarization for these states, the matrix
element becomes an expectation value and thus acquires a probability
interpretation akin to the usual parton densities.

The wave packets (\ref{impact-states}) involve proton momenta which
are not along the $z$-axis, and the spin states for this case have to
be chosen with some care.  It is useful to take states of definite
light-cone helicity~\cite{Soper:1972xc}.  A proton state of momentum
$p$ with positive (negative) light-cone helicity is transformed to a
proton state at rest with spin along the positive (negative) $z$-axis
by a Lorentz transformation $\mathcal{L}(p)$, which is the combination
of a transverse and a longitudinal boost (see Sect.~3.5.1 of
\cite{Diehl:2003ny} for a brief summary).  The light-cone helicity of
a state is invariant under boosts along the $z$-axis, and for large
$p^+$ light-cone helicity coincides with the usual helicity up to
effects of order $m/p^+$.  The superposition $(\, |+\rangle +
e^{i\varphi} |-\rangle \, )/ \sqrt{2}$ of states with positive and
negative light-cone helicity is called a state of definite
transversity, which can be seen as the light-cone analog of definite
transverse polarization.  According to what we just discussed,
$\mathcal{L}(p)$ indeed transforms this state to a state at rest whose
spin vector is given by $\tvec{S} = (\cos\varphi, \sin\varphi)$ and
$S^z=0$.  A state with both longitudinal and transverse polarization
can be written as
\begin{equation}
|\Lambda,\tvec{S}\rangle=
\cos(\half\vartheta)\, |+\rangle +
\sin(\half\vartheta)\, e^{i\varphi} |-\rangle 
\label{long-trans-states}
\end{equation}
and is transformed by $\mathcal{L}(p)$ to a state at rest with spin
vector given by $\tvec{S} = (\sin\vartheta\, \cos\varphi,
\sin\vartheta\, \sin\varphi)$ and $S^z = \cos\vartheta$.  We will
therefore use $\tvec{S}$ and $\Lambda = S^z$ to characterize these
states.  Combining them to wave packets (\ref{impact-states}) we
finally obtain states suitable for interpreting the matrix elements
$F(x,\tvec{b})$, $\tilde{F}(x,\tvec{b})$ and $F_T^j(x,\tvec{b})$.  For
ease of language we will call $\tvec{S}$ and $\Lambda$ the transverse
and longitudinal polarization of the proton.  In the following it will
be important that they respectively transform like a usual spin vector
and usual helicity under rotations in the $x$-$y$ plane and under
parity or time reversal.

For quarks and antiquarks we consider light-cone helicity states as
well.  Note that the quark operators in (\ref{impact-mat-elements})
are at definite transverse position and thus correspond to integrals
over the quark or antiquark transverse momentum.  Quarks with
light-cone helicity $\lambda= \pm 1$ are projected out by the operator
$\half \bar{q}\, \gamma^+ [1 + \lambda \gamma_5]\, q$.  Evaluating the
proton spinor products in (\ref{gpd-def}) for the states
(\ref{long-trans-states}) and Fourier transforming the result, we
obtain the density of quarks with light-cone helicity $\lambda$,
light-cone momentum fraction $x$ and transverse distance $\tvec{b}$
from the proton center as
\begin{eqnarray}
  \label{long-distr}
\frac{1}{2} \Big[ F(x,\tvec{b}) 
                  + \lambda \tilde{F}(x,\tvec{b}) \,\Big] 
 &=& \frac{1}{2} \Bigg[ H(x,\tvec{b}^2) 
  - S^i \epsilon^{ij} b^j \frac{1}{m}\, 
       \frac{\partial}{\partial \tvec{b}{}^2} E(x,\tvec{b}^2)
  + \lambda \Lambda \tilde{H}(x,\tvec{b}^2) \,\Bigg] 
\end{eqnarray}
for $x>0$, where repeated Roman indices are to be summed over.  For
$x<0$ the density of antiquarks with light-cone helicity $\lambda$,
light-cone momentum fraction $-x$ and transverse position $\tvec{b}$
is given by $\smash{\half} [ - F(x,\tvec{b}) + \lambda
\tilde{F}(x,\tvec{b}) \,]$.  It readily follows from the
transformation properties of $\bar{q} \gamma^+ q$ and $\bar{q}
\gamma^+ \gamma_5\, q$ under charge conjugation that in going from
quark to antiquark densities one has to change the sign of $F$ but not
of $\tilde{F}$.  The result (\ref{long-distr}) is well-known and for
instance discussed in \cite{Burkardt:2002hr}.  The term with
$\tilde{H}$ reflects the difference in density of quarks with helicity
equal or opposite to the proton helicity.  More remarkably, the term
with $E$ describes a sideways shift in the unpolarized quark density
when the proton is transversely polarized.

We now discuss transverse quark and antiquark polarization, which in
analogy to our above discussion we define as the superposition $(\,
|+\rangle + e^{i\chi} |-\rangle \, )/ \sqrt{2}$ of positive and
negative light-cone helicities, with a transverse spin vector
$\tvec{s} = (\cos\chi, \sin\chi)$.  Quarks with transverse
polarization $\tvec{s}$ are projected out by the operator $\half
\bar{q} \gamma^+ [1 + (\tvec{s}\tvec{\gamma}) \gamma_5]\, q = \half
\bar{q}\, [\gamma^+ - s^j i \sigma^{+j} \gamma_5 ]\, q$, and their
density is
\begin{eqnarray}
  \label{trans-distr}
\frac{1}{2} \Big[ F + s^i F_T^i \,\Big]
 &=& \frac{1}{2} \Bigg[ H 
  - S^i \epsilon^{ij} b^j \frac{1}{m}\, E'
  - s^i \epsilon^{ij} b^j \frac{1}{m}
    \Big( E_T' + 2 \tilde{H}_T' \Big)
\nonumber \\
 && {}+ s^i S^i \Big( H_T - \frac{1}{4m^2}\, \Delta_b \tilde{H}_T \Big)
  + s^i (2 b^i b^j - b^2 \delta^{ij}) S^j \frac{1}{m^2} 
    \tilde{H}_T'' \,\Bigg]
\end{eqnarray}
for $x>0$.  The density of antiquarks with transverse polarization
$\tvec{s}$ and light-cone momentum fraction $-x$ is given by $-
\smash{\half} [ F(x,\tvec{b}) + s^i F_T^i(x,\tvec{b}) \,]$.  Here and
in the following it is understood that when nothing else is indicated,
the matrix elements $F$, $\tilde{F}$, $\smash{F_T^j}$ and the
distributions $H$, $E$, $\tilde{H}$, $H_T$ etc.\ are functions of $x$
and $\tvec{b}$ as given in (\ref{impact-mat-elements}) and
(\ref{impact-gpds}).  We write $b= \sqrt{\tvec{b}{}^2}$ so that $b^2 =
\tvec{b}{}^2$ and $\partial /\partial b^2 =\partial /\partial
\tvec{b}{}^2$, and we use the shorthand
\begin{equation}
  \label{deriv-def}
f' = \frac{\partial}{\partial b^2}\, f ,
\qquad\qquad
f''= \Big( \frac{\partial}{\partial b^2} \Big)^2 f
\end{equation}
for the derivatives and
\begin{equation}
  \label{laplace}
\Delta_b f
= \frac{\partial}{\partial b^i}\, \frac{\partial}{\partial b^i}\, f
= 4\, \frac{\partial}{\partial b^2}
    \Big( b^2 \frac{\partial}{\partial b^2} \Big) f 
\end{equation}
for the two-dimensional Laplace operator acting on functions $f$ that
depend on $\tvec{b}$ only via its square.  In (\ref{trans-distr}) we
have further introduced the two-dimensional antisymmetric tensor
$\epsilon^{ij}$ with $\epsilon^{12} = -\epsilon^{21} = 1$ and
$\epsilon^{11} = \epsilon^{22} = 0$.

\begin{figure}
\begin{center}
\leavevmode
\includegraphics[height=15cm,angle=90,clip=true]{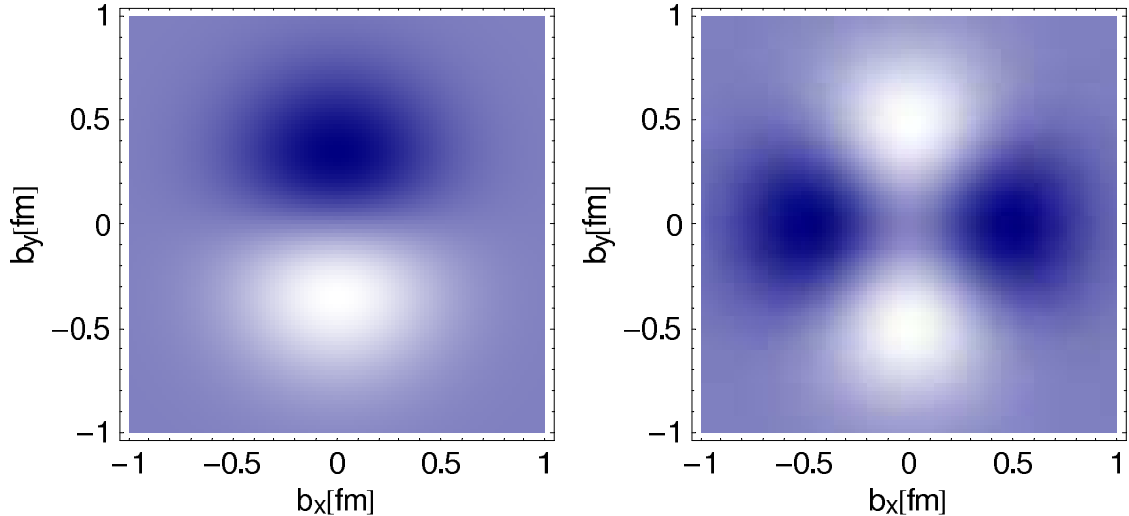}
\end{center}
\caption{\label{fig:density-terms} Density plots in the impact
  parameter plane for the functions $b \exp[-b^2 /b_0^2] \sin\phi$
  (left) and $b^2 \exp[-b^2 /b_0^2] \cos(2\phi)$ (right), with $b_0 =
  0.5$~fm.  These functions illustrate the terms in the quark density
  (\protect\ref{trans-distr}) which break rotational symmetry, as
  explained after (\protect\ref{density-terms}).  Dark areas represent
  high densities.}
\end{figure}

The term with $E_T' + 2 \tilde{H}_T'$ in (\ref{trans-distr}) describes
a sideways shift in the distribution of transversely polarized quarks
in an unpolarized proton, whereas the last two terms in that
expression reflect a correlation in the quark density between the
transverse polarizations of quark and proton.  The structures which
break rotational symmetry in the density (\ref{trans-distr}) are
\begin{eqnarray}
  \label{density-terms}
S^i \epsilon^{ij} b^j &=& b \sin\phi ,
\nonumber \\
s^i \epsilon^{ij} b^j &=& b \sin(\phi - \chi) ,
\nonumber \\
s^i (2 b^i b^j - b^2 \delta^{ij}) S^j &=& b^2 \cos(\chi - 2\phi) ,
\end{eqnarray}
where we have parameterized $\tvec{b} = b\, (\cos\phi, \sin\phi)$ and
taken $\tvec{S} = (1,0)$ for simplicity.  For illustration we show in
Fig.~\ref{fig:density-terms} density plots for the functions $b
\exp[-b^2 /b_0^2] \sin\phi$ and $b^2 \exp[-b^2 /b_0^2] \cos(2\phi)$,
where the exponentials are taken to mimic the impact parameter
dependence of the relevant parton distributions.

Integrating the densities (\ref{long-distr}) and (\ref{trans-distr})
over all impact parameters one obtains generalized parton
distributions in momentum space at $t=0$, namely
\begin{eqnarray}
  \label{b-int-1a}
\int d^2\tvec{b}\,
  \Big[ F(x,\tvec{b}) + \lambda \tilde{F}(x,\tvec{b}) \Big]
\:=\: H(x,0,0) + \lambda \Lambda \tilde{H}(x,0,0)
 &=&  f_1(x) + \lambda \Lambda\, g_1(x) \; , 
\\
\label{b-int-1b}
\int d^2\tvec{b}\,
  \Big[ F(x,\tvec{b}) + s^i F_T^i(x,\tvec{b}) \Big]
\:=\: H(x,0,0) + s^i S^i\, H_T(x,0,0)
 &=&  f_1(x) + s^i S^i\, h_1(x) \; .
\end{eqnarray}
Here we have used that GPDs in the forward limit $\xi=0$ and $t=0$
reduce to the usual parton densities, namely
\begin{equation}
  \label{forward-limits}
H(x,0,0) = f_1(x) , \qquad\qquad
\tilde{H}(x,0,0) = g_1(x) , \qquad\qquad 
H_T(x,0,0) = h_1(x)
\end{equation}
for $x>0$, where $f_1$ denotes the unpolarized quark distribution,
$g_1$ the quark helicity distribution, and $h_1$ the quark
transversity distribution (another common notation is $f_1 = q$,
$g_1=\Delta q$ and $h_1=\delta q$).  Corresponding relations involving
antiquark distributions hold for $x<0$.  Weighting the impact
parameter distributions with $b^2$ before integration, one obtains
derivatives at $t=0$,
\begin{eqnarray}
  \label{b-int-2a}
\int d^2\tvec{b}\; b^2 \Big( F + \lambda \tilde{F} \Big) 
 &=& \Bigg[ 4 \frac{\partial}{\partial t} 
       \Big( H + \lambda \Lambda \tilde{H} \Big) \Bigg]_{t=0} \; ,
\\
  \label{b-int-2b}
\int d^2\tvec{b}\; b^2 \Big( F + s^i F_T^i \Big)
 &=& \Bigg[ 4 \frac{\partial}{\partial t} \Big(  H + s^i S^i
       \Big[ H_T -  \frac{t}{4m^2} \tilde{H}_T \Big] \Big)
       \Bigg]_{t=0} \, ,
\end{eqnarray}
where we use the subscript $t=0$ to indicate that the GPDs are taken
in momentum space as in~(\ref{gpd-def}), with $\xi=0$ as always in
this and the next section.  The ratio of the integrals in
(\ref{b-int-2a}) and (\ref{b-int-1a}) or in (\ref{b-int-2b}) and
(\ref{b-int-1b}) thus gives the average squared impact parameter $b^2$
of quarks with given polarization and plus-momentum fraction.  The
average sideways shift in the impact parameter distribution due to the
transverse polarization of either the quark or the proton is obtained
from
\begin{equation}
  \label{b-int-3}
\int d^2\tvec{b}\; b^j\, \Big( F + s^i F_T^i \Big)
 = \Bigg[ \frac{1}{2m} \Big( S^i \epsilon^{ij} E
   + s^i \epsilon^{ij} ( E_T + 2 \tilde{H}_T ) \Big) \Bigg]_{t=0}
\end{equation}
normalized to the integral in (\ref{b-int-1b}).  This shift is more
involved than in the case of longitudinally polarized or unpolarized
quarks,
\begin{equation}
  \label{b-int-3b}
\int d^2\tvec{b}\; b^j\, \Big( F +  \lambda \tilde{F}  \Big)
 = \int d^2\tvec{b}\; b^j\, F
 = \Bigg[ \frac{1}{2m} S^i \epsilon^{ij} E \,\Bigg]_{t=0}\; ,
\end{equation}
which has been discussed in some detail in \cite{Burkardt:2002hr}.
Finally, the average distortion of the impact parameter density due to
the last term in (\ref{trans-distr}) is characterized by
\begin{equation}
  \label{b-int-4}
\int d^2\tvec{b}\; (2 b^j b^k - b^2 \delta^{jk})\,
     \Big( F + s^i F_T^i \Big)
 = \Bigg[ \frac{1}{m^2}\, (s^j S^k + S^j s^k - s^i S^i\, \delta^{jk})\,
   \tilde{H}_T \,\Bigg]_{t=0} \, .
\end{equation}

We note in (\ref{long-distr}) and (\ref{trans-distr}) that there is no
polarization effect in the impact parameter distributions for
longitudinally polarized quarks in a transversely polarized proton and
vice versa.  This is because the only structures describing such
effects which are allowed by parity conservation are $\lambda S^i b^i$
or $\Lambda s^i b^i$.  They are odd under time reversal and hence
forbidden.  This corresponds to the fact that the generalized
distributions $\tilde{E}$ and $\tilde{E}_T$ in (\ref{gpd-def}) do not
appear in the matrix elements at $\xi=0$, the former because it is
multiplied with $\Delta^+ = -2\xi P^+$ in its definition and the
latter because it is an odd function of $\xi$ \cite{Diehl:2001pm}.

It is instructive to compare our impact parameter densities with the
densities for quarks of definite light-cone momentum fraction $x$ and
transverse momentum $\tvec{k}$, which play a prominent role in the
description of spin asymmetries in a variety of hard processes.  They
can be defined from the correlation function
\begin{equation}
  \label{correl-fct}
\Phi_{\alpha\beta}(x,\tvec{k}) = 
\int \frac{dz^-}{4\pi} \frac{d^2\tvec{z}}{(2\pi)^2}\;
    e^{i x p^+ z^-} e^{-i \svec{k} \svec{z}}\,
    \langle p|\, \bar{q}_\beta(-\half z)\, W[-\half z, \half z]\;
    q_\alpha(\half z) |\, p\rangle \,\Big|_{z^+ =0} \; ,
\end{equation}
where it is understood that the proton states have zero transverse
momentum.  The Wilson line $W$ has recently been recognized as
essential in the definition, since different physical processes
require different paths leading from $-\half z$ to $\half z$ and can
actually give different correlation functions (see e.g.\
\cite{Boer:2003cm} and references therein).  Physically, the gluons
resummed in the Wilson lines describe interactions of spectator
partons in the process where the correlation function appears, and the
corresponding parton distributions describe the density of quarks or
antiquarks in the presence of these interactions.  This subtlety did
not appear in the generalized parton distributions (\ref{gpd-def}) and
their impact parameter analogs, because there the quark field and its
conjugate are separated by a light-like distance and the relevant
Wilson line just runs along the light-cone between $-\half z$ and
$\half z$.  Projecting out densities for quarks of definite
longitudinal or transverse polarization from (\ref{correl-fct}), one
obtains \cite{Boglione:1999pz}
\begin{eqnarray}
  \label{piet-distr}
\frac{1}{2} \mathrm{Tr} \Big[
  ( \gamma^+ + \lambda \gamma^+ \gamma_5 )\, \Phi \Big]
 &=& \frac{1}{2} \Bigg[ f_1^{\phantom{\perp}\!\!}
   + S^i \epsilon^{ij} k^j \frac{1}{m}\, f_{1T}^\perp
   + \lambda \Lambda\, g_1^{\phantom{\perp\!\!}}
   + \lambda\, S^i k^i \frac{1}{m}\, g_{1T}^{\phantom{\perp\!\!}} 
  \Bigg] \, ,
\nonumber \\
\frac{1}{2} \mathrm{Tr} \Big[
  ( \gamma^+ - s^j i\sigma^{+j} \gamma_5 )\, \Phi \Big]
 &=& \frac{1}{2} \Bigg[ f_1^{\phantom{\perp\!\!}}
   + S^i \epsilon^{ij} k^j \frac{1}{m}\, f_{1T}^\perp
   + s^i \epsilon^{ij} k^j \frac{1}{m}\, h_{1}^\perp
   + s^i S^i h_1^{\phantom{\perp\!\!}}
\nonumber \\
 && \hspace{0.7em}
 {}+ s^i (2 k^i k^j - \tvec{k}^2 \delta^{ij}) S^j 
       \frac{1}{2m^2}\, h_{1T}^\perp
   +  \Lambda\, s^i k^i \frac{1}{m}\, h_{1L}^\perp \Bigg] \, ,
\end{eqnarray}
where we have used the notation of Boer, Mulders, Tangerman
\cite{Mulders:1995dh,Boer:1997nt} for the distribution functions,
which depend on $x$ and $\tvec{k}^2$.  Integrating over $\tvec{k}$ one
recovers the distributions $f_1(x) = \int d^2\tvec{k}\,
f_1(x,\tvec{k}^2)$, $g_1(x) = \int d^2\tvec{k}\, g_1(x,\tvec{k}^2)$
and $h_1(x) = \int d^2\tvec{k}\, h_1(x,\tvec{k}^2)$ we already
encountered in (\ref{forward-limits}).  The tensor structures in
(\ref{piet-distr}) are analogs of those in (\ref{long-distr}) and
(\ref{trans-distr}), with $\tvec{k}$ taking the role of $\tvec{b}$.
The corresponding analogy between transverse momentum dependent and
impact parameter dependent distributions reads
\begin{eqnarray}
  \label{dictionary}
&& 
f_1^{\phantom{\perp}} \leftrightarrow H , \hspace{9.3em}
f_{1T}^\perp          \leftrightarrow {}- E' , \hspace{6.6em}
g_1^{\phantom{\perp}} \leftrightarrow \tilde{H} ,
\nonumber \\[0.2em]
&& 
h_1^{\phantom{\perp}} \leftrightarrow
    H_T - \Delta_b \tilde{H}_T /(4m^2) \, , \qquad
h_1^\perp      \leftrightarrow {}- (E_T' + 2\tilde{H}_T') \, , \qquad
h_{1T}^\perp   \leftrightarrow 2 \tilde{H}_T'' \, .
\end{eqnarray}
The impact parameter distributions which would correspond to
$g_{1T}^{\phantom{\perp}}$ and $h_{1L}^\perp$ vanish because of time
invariance, as discussed above.  Notice that the momentum $\tvec{k}$
changes sign under time reversal, whereas the position vector
$\tvec{b}$ does not.  The structures $\lambda\, S^i k^i$ and
$\Lambda\, s^i k^i$ describing polarization effects for longitudinally
polarized quarks in a transversely polarized proton and vice versa are
hence time reversal invariant.  On the other hand, both $S^i
\epsilon^{ij} k^j$ and $s^i \epsilon^{ij} k^j$ are odd under time
reversal.  The corresponding distributions $f_{1T}^\perp$ and
$h_{1}^\perp$ (which respectively are the Sivers and Boer-Mulders
functions) are however not constrained to be zero by time reversal
invariance.  This is because the Wilson line in the correlation
function for relevant processes like semi-inclusive deep inelastic
scattering or Drell-Yan pair production have paths that are \emph{not}
invariant under time reversal, contrary to the paths appearing in the
impact parameter distributions discussed above.  Time reversal thus
connects transverse momentum dependent distributions with different
Wilson lines, but does not constrain them to be zero
\cite{Collins:2002kn}.  We finally note that, beyond the formal
correspondence of the functions $E(x,\tvec{b})$ in (\ref{trans-distr})
and $f_{1T}^\perp(x,\tvec{k})$ in (\ref{piet-distr}), a deep dynamical
connection between them has recently been proposed in
\cite{Burkardt:2003uw,Burkardt:2003je}.

We have seen in (\ref{b-int-1a}) to (\ref{b-int-4}) how the impact
parameter distributions can be reduced to distributions only depending
on the momentum fraction $x$ by taking appropriate integrals over
$\tvec{b}$.  Conversely, one obtains distributions only depending on
$\tvec{b}$ by integrating over $x$.  Taking Mellin moments in~$x$, we
obtain expectation values of the well-known local twist-two operators
in proton states localized at zero impact parameter,
\begin{eqnarray}
  \label{mellin-moments}
(p^+)^{n} \int_{-1}^1 dx\, x^{n-1} F(x,\tvec{b}) &=& 
  \frac{1}{2}\, \mathcal{N}^{-1}\,
  \Big\langle p^+,\tvec{0} \,\Big|\, 
     \bar{q} \gamma^{+}\, (i\lrD^+)^{n-1} q 
  \Big|\, p^+,\tvec{0}\Big\rangle ,
\nonumber \\
(p^+)^{n} \int_{-1}^1 dx\, x^{n-1} \tilde{F}(x,\tvec{b}) &=&
  \frac{1}{2}\, \mathcal{N}^{-1}\,
  \Big\langle p^+,\tvec{0} \,\Big|\, 
     \bar{q} \gamma^{+} \gamma_5\, (i\lrD^+)^{n-1} q
  \Big|\, p^+,\tvec{0}\Big\rangle ,
\nonumber \\
(p^+)^{n} \int_{-1}^1 dx\, x^{n-1} F_T^j(x,\tvec{b}) &=&
  {}- \frac{i}{2}\, \mathcal{N}^{-1}\,
  \Big\langle p^+,\tvec{0} \,\Big|\, 
     \bar{q} \sigma^{+j} \gamma_5\, (i\lrD^+)^{n-1} q 
  \Big|\, p^+,\tvec{0}\Big\rangle ,
\end{eqnarray}
where $\lrD^\mu = \half (\rD^\mu - \lD^\mu) = \half
(\hspace{0.2pt}\rpartial^\mu - \lpartial^\mu) - ig A^\mu$, and all
field operators are to be taken at position $z$ with $z^+ = z^- = 0$
and $\tvec{z} = \tvec{b}$.  To obtain matrix elements of local
operators, one has to integrate over both positive and negative $x$
and hence not only combines the information from different momentum
fractions but also from quarks and antiquarks.  According to the
charge conjugation properties we discussed after (\ref{long-distr})
and (\ref{trans-distr}), moments with odd $n$ in
(\ref{mellin-moments}) correspond to the sum of quark and antiquark
densities for $\tilde{F}$ and to their difference for $F$ and $F_T^j$,
whereas for moments with even $n$ the situation is reversed.  The
lowest $x$ moments of $F$ and $\half [ F + s^i F_T^i ]$ hence describe
the transverse distribution of unpolarized and transversely polarized
quarks minus antiquarks in the proton, respectively.  Higher $x$
moments give the transverse distributions of quarks plus or minus
antiquarks weighted with a power of their plus-momentum fraction.  In
contrast, the $x$ moments of $\half [F + \lambda \tilde{F}]$ describe
the transverse distribution of quarks plus or minus antiquarks with
\emph{chirality} $\lambda$ (i.e.\ quarks with helicity $\lambda$ and
antiquarks with helicity $-\lambda$).  A two-dimensional Fourier
transform turns the expectation values (\ref{mellin-moments}) into
matrix elements for proton states of definite momenta, which are
parameterized by form factors depending on the squared momentum
transfer $t$.  These form factors can be evaluated in lattice QCD
since the corresponding operators are local and thus allow
continuation into Euclidean space.  It is amusing that after a Fourier
transform they become quantities (\ref{mellin-moments}) whose physical
interpretation is naturally given in a light-cone framework.

Note that our interpretation of form factors differs from the
well-known interpretation due to Sachs~\cite{Sachs:1962aa}, where
their \emph{three-dimensional} Fourier transforms yield densities in
a static proton state.  That framework has been extended to
generalized parton distributions as functions of $x$, $\xi$ and $t$
in~\cite{Belitsky:2003nz}.  It is limited by ambiguities due to the
impossibility to localize a particle more accurately than within its
Compton wavelength.  Localization in only two dimensions is not
affected by this limitation, and the wave packets
(\ref{impact-states}) are indeed eigenstates of a two-dimensional
position operator~\cite{Soper:1972xc}.  The price to pay for this is
the loss of manifest three-dimensional rotation invariance in the
light-cone framework.  In return, the mixed representation of position
space in two dimensions and plus-momentum in the third allows one to
boost to a frame where the proton moves fast, which is a natural frame
for the physical interpretation of quark and antiquark degrees of
freedom.

\begin{figure}
\begin{center}
\leavevmode
\includegraphics[height=15cm,angle=90,clip=true]{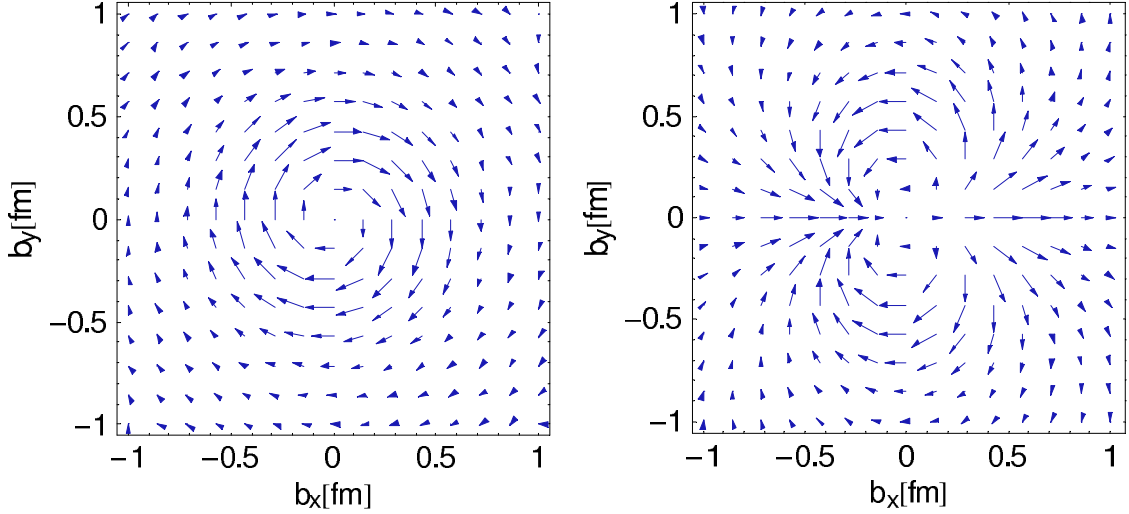}
\end{center}
\caption{\label{fig:fieldline-terms} The vector fields $\epsilon^{ij}
b^j \exp[-b^2 /b_0^2]$ (left) and $(2 b^i b^j - b^2 \delta^{ij}) S^j
\exp[-b^2 /b_0^2]$ (right) with $S^i$ along the $x$-axis and $b_0 =
0.5$~fm.  They illustrate the form of two terms in the decomposition
of $F_T^i(x,\tvec{b})$, which describes the transverse polarization of
quarks in the impact parameter plane.  The third term in the
decomposition is a field parallel to $S^i$.}
\end{figure}

So far we have interpreted $\half [F + s^i F_T^i ]$ in
(\ref{trans-distr}) as the density of quarks with a given transverse
polarization $\tvec{s}$.  The vector field
$\tvec{F}_{\!T}(x,\tvec{b})$ gives the direction in which the
transverse polarization of quarks is largest, and its size
$|\tvec{F}_{\!T}(x,\tvec{b})|$ is the difference of densities with
quarks polarized along or opposite to this direction.  According to
(\ref{trans-distr}) we can write $F_T^i$ as the superposition of three
terms, given by functions of $\tvec{b}^2$ times the vectors $S^i$,
$\epsilon^{ij} b^j$ and $(2 b^i b^j - b^2 \delta^{ij}) S^j$.  The
field lines of the term with $\epsilon^{ij} b^j$ are circles around
the origin, and those of the term with $(2 b^i b^j - b^2 \delta^{ij})
S^j$ are circles going through the origin with a tangent along the
proton polarization $S^i$, as illustrated in
Fig.~\ref{fig:fieldline-terms}.

The divergence and the curl of the field $F_T^i(x,\tvec{b})$
respectively are
\begin{eqnarray}
  \label{div-curl}
\frac{\partial}{\partial b^i} F_T^i
  &=& 2 S^i b^i H_T' \, ,
\nonumber \\
\frac{\partial}{\partial b^i} \epsilon^{ij} F_T^j 
  &=& \frac{1}{2m} \Delta_b \Big( E_T + 2 \tilde{H}_T \Big)
   -2 S^i \epsilon^{ij} b^j
      \frac{\partial}{\partial b^2} 
      \Big( H_T - \frac{1}{2m^2} \Delta_b \tilde{H}_T \Big) \, .
\end{eqnarray}
They can be rewritten as matrix elements of operators which are total
derivatives.  This is readily seen in Mellin space, where we have
\begin{eqnarray}
  \label{mellin-derivative}
(p^+)^{n} \int_{-1}^1 dx\, x^{n-1}
  \frac{\partial}{\partial b^i} F_T^i(x,\tvec{b})  
  &=& {}- \frac{i}{2}\, \mathcal{N}^{-1}\,
  \Big\langle p^+,\tvec{0} \,\Big|\, \partial_\mu 
     \Big[ \bar{q} \sigma^{+\mu} \gamma_5\, (i\lrD^+)^{n-1} q \Big]\,
  \Big|\, p^+,\tvec{0}\Big\rangle \, ,
\nonumber \\
(p^+)^{n} \int_{-1}^1 dx\, x^{n-1}
  \frac{\partial}{\partial b^i} \epsilon^{ij} F_T^j(x,\tvec{b})  
  &=& \frac{1}{2}\, \mathcal{N}^{-1}\,
  \Big\langle p^+,\tvec{0} \,\Big|\, \partial_\mu
     \Big[ \bar{q} \sigma^{+\mu} (i\lrD^+)^{n-1} q \Big]\,
  \Big|\, p^+,\tvec{0}\Big\rangle \, ,
\end{eqnarray}
with all field operators evaluated at $z^+ = z^- = 0$ and $\tvec{z} =
\tvec{b}$.  To obtain (\ref{mellin-derivative}) we have used the
representation (\ref{mellin-moments}) and the fact that the first term
in $\partial^+ [\bar{q} \sigma^{+-} \ldots q] + \partial_i [\bar{q}
\sigma^{+i} \ldots q] = \partial_\mu [\bar{q} \sigma^{+\mu} \ldots q]$
vanishes when taking a matrix element between states of equal
plus-momentum.  The operators in (\ref{mellin-derivative}) can be
rewritten using the equations of motion as we will discuss in
Sect.~\ref{sec:eom}.

The Mellin moments of the impact parameter distributions $H(x,b^2)$,
$E(x,b^2)$ etc.\ are Fourier transforms of form factors in momentum
space, which are denoted by
\begin{eqnarray}
A_{n0}(t) &=& \int_{-1}^1 dx\, x^{n-1} H(x,0,t) , \qquad\hspace{1.05em}
B_{n0}(t) \:=\: \int_{-1}^1 dx\, x^{n-1} E(x,0,t) ,
\nonumber \\
A_{T n0}(t) &=& \int_{-1}^1 dx\, x^{n-1} H_T(x,0,t) , \qquad
B_{T n0}(t) \:=\: \int_{-1}^1 dx\, x^{n-1} E_T(x,0,t) , 
\nonumber \\
\tilde{A}_{T n0}(t) &=& \int_{-1}^1 dx\, x^{n-1} \tilde{H}_T(x,0,t)
\end{eqnarray}
in a standard notation (see Sect.~\ref{sec:eom}).  These form factors
can be calculated in lattice QCD
\cite{Gockeler:2003jf,Hagler:2003jd,Gockeler:2005aw}, where it has
become customary to fit them to a dipole form.  For reasons that will
be clear shortly, let us consider the more general power-law ansatz
\begin{equation}
A(t) = \frac{A(0)}{(1 - {t/\mm^2}
        )^{\hspace{1pt} p}_{\phantom{d}} \rule{0em}{1em}} \: ,
\label{dipole}
\end{equation}
where the power $p$ and the mass $\mm$ are free parameters for a given
form factor $A(t)$.  Note that in the limit $p\to \infty$ at fixed
$\mm^2 /p$ this ansatz gives an exponential in $t$.  The Fourier
transformation of (\ref{dipole}) to impact parameter space leads to
the modified Bessel function,
\begin{equation}
A(b^2) = C\; (\mm b)^{p-1} K_{p-1}(\mm b) \, , \qquad\qquad
C = \frac{\mm^2}{2^p \pi \Gamma(p)}\, A(0) \, ,
\end{equation}
and the derivatives defined in (\ref{deriv-def}) and (\ref{laplace})
are 
\begin{eqnarray}
A'(b^2)  &=& - \half C \mm^2\, (\mm b)^{p-2} K_{p-2}(\mm b) \, ,
\qquad\qquad
A''(b^2) = {\textstyle\frac{1}{4}} C \mm^4\,
           (\mm b)^{p-3} K_{p-3}(\mm b) \, ,
\nonumber \\[0.3em]
\Delta_b A(b^2) &=& - C \mm^2\, (\mm b)^{p-2}
  \Big[ 2 K_{p-2}(\mm b) - \mm b\, K_{p-3}(\mm b) \,\Big] \,.
\label{dipole-deriv}
\end{eqnarray}
A parameterization of the type (\ref{dipole}) is in the first instance
only valid in the $t$ range where the form factor has been fitted.  In
particular, lattice computations have an upper limit
$|t|_\mathrm{max}$ on the squared momentum transfer given by the
lattice parameters.  This corresponds to a limited resolution of order
$(|t|_\mathrm{max})^{-1/2}$ on the impact parameter
\cite{Diehl:2002he}.  Furthermore, results obtained on a finite
lattice cannot give direct information on the behavior of quark
densities at impact parameters much larger than the lattice size.
Nevertheless, one may want to require that a parameterization leads at
least to a physically plausible behavior of the impact parameter
density at small and large $b$.  To analyze this behavior we need the
relations
\begin{eqnarray}
K_0(z) &\stackdown{\sim}{z\to 0}& \log\frac{2}{z} , \qquad\qquad
K_p(z) \:\:\:\stackdown{\sim}{z\to 0}\:\:\:
       \frac{2^{p-1} \Gamma(p)}{z^{p}}
\qquad \mbox{for $p>0$} ,
\nonumber \\
K_p(z) &\stackdown{\sim}{z\to\infty}& e^{-z} \sqrt{\frac{\pi}{2z}}
\end{eqnarray}
and $K_{-p}(z) = K_p(z)$.  At large $b$, each term in the Mellin
moments of the densities (\ref{long-distr}) and (\ref{trans-distr})
then falls off like $(\mm b)^{p-3/2}\, e^{-\mm b}$.  For the limit
$b\to 0$ it seems reasonable to require a regular behavior of the
impact parameter density, which implies that no term should diverge
and that $b^j E'$, $b^j (E_T' + 2\tilde{H}_T')$ and $(2 b^i b^j - b^2
\delta^{ij}) \tilde{H}_T''$ should vanish at $b=0$, since they have a
nontrivial dependence on the azimuthal angle $\phi$.  This restricts
$p$ in the parameterization of moments to $p>1$ for $H$, $\tilde{H}$
and $H_T$, to $p>3/2$ for $E$ and $E_T$, and to $p>2$ for
$\tilde{H}_T$.  The terms with $H$, $\tilde{H}$, $H_T$ and $\Delta_b
\tilde{H}_T$ in the moments of (\ref{long-distr}) and
(\ref{trans-distr}) then all take finite values at $b=0$.  In momentum
space these restrictions are tantamount to requiring that the Mellin
moments of $F(x,0,t)$, $\tilde{F}(x,0,t)$, $F_T^j(x,0,t)$ decrease
faster than $1/t$ for $t\to -\infty$, as is readily seen when the
proton spinor products in (\ref{gpd-def}) are evaluated
\cite{Diehl:2001pm}.  Note in particular that a dipole ansatz with
$p=2$ for $\tilde{A}_{T n0}(t)$ gives only a $1/t$ falloff in the
$n$th moment of $F_T^i(x,0,t)$ and a corresponding logarithmic
divergence at $b=0$ in the $n$th moment of $F_T^i(x,\tvec{b})$.

To illustrate how the transverse spin density in the proton may look
like, we focus now on the first moment $\half \int_{-1}^1 dx\,
[F(x,\tvec{b}) + s^i F_T^i(x,\tvec{b})]$, which gives the difference
of impact parameter densities for quarks and antiquarks (for ease of
language we will simply speak of quarks in the following).  As a
numerical example we take a parameterization (\ref{dipole}) with $p=2$
for $A_{10}(t)$, $B_{10}(t)$, $A_{T 10}(t)$ and $p=3$ for $B_{T
10}(t)$, $\tilde{A}_{T 10}(t)$.  We set the mass parameters $\mm$ to
$1$~GeV for $A_{10}(t)$, $B_{10}(t)$ and to $1.5$~GeV for the three
other form factors, and take
\begin{equation}
  \label{zero-values}
A_{10}(0) = 2, \qquad
B_{10}(0) = 3, \qquad
A_{T 10}(0) = 1, \qquad
B_{T 10}(0) = 6, \qquad
\tilde{A}_{T 10}(0) = -1
\end{equation}
for their values at $t=0$.  This set of parameterizations is a rough
approximation of preliminary results from lattice calculations
\cite{QCDSF:2005nn} for the first moments of generalized $u$-quark
distributions (where $|t|$ goes up to about $3.5\, \mbox{GeV}^2
\approx (0.1\, \mbox{fm})^{-2}$ and lattice sizes are between $1.5$
and $2.2$~fm).  We stress that it is meant to be indicative and
\emph{not} a precise representation of those results.  We note that
$A_{10}(0) = 2$ correctly gives the number of valence $u$-quarks in
the proton, whereas $B_{10}(0) = 3$ is too large compared with the
value $1.67$ one obtains from the measured magnetic moments of proton
and neutron (recall that $B_{10}(t)$ is the relevant quark flavor
contribution to the electromagnetic Pauli form factor).

\begin{figure}
\begin{center}
\leavevmode
\includegraphics[height=15cm,angle=90,clip=true]{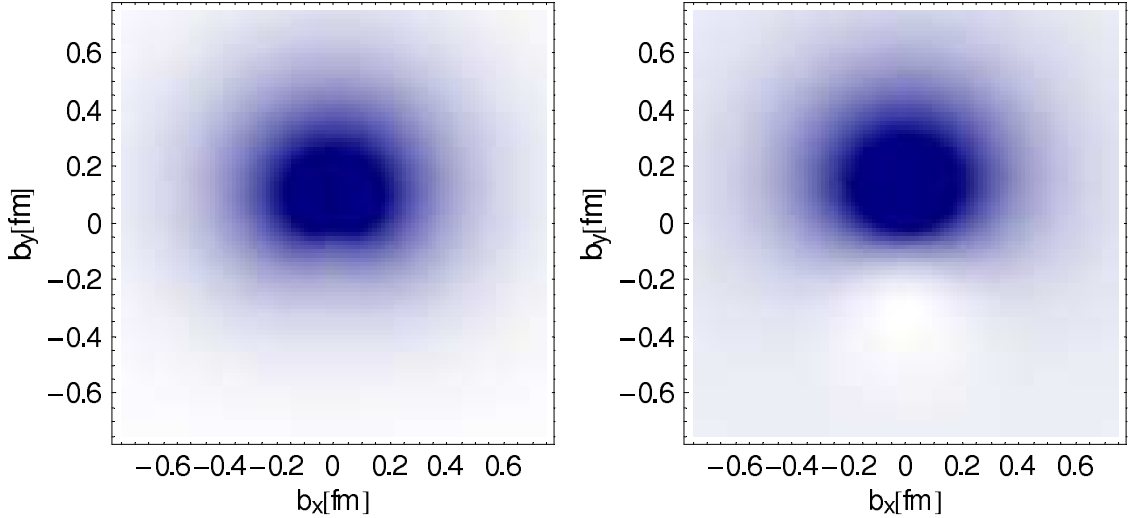}
\end{center}
\caption{\label{fig:impact-density-0} Left: Illustration of the first
  moment $\int_{-1}^1 dx\, F(x,\tvec{b})$ of the impact parameter
  density for unpolarized $u$-quarks in a proton with transverse spin
  vector $\tvec{S}=(1,0)$.  Right: The same for the first moment
  $\half \int_{-1}^1 dx\, [F(x,\tvec{b})+s^i F_T^i(x,\tvec{b})]$ of
  the distribution of $u$-quarks with transverse spin vector
  $\tvec{s}=(1,0)$ in an unpolarized proton.  Dark areas represent the
  highest and light areas the lowest values of the density.  Further
  explanation is given in the text.}
\end{figure}

\begin{figure}
\begin{center}
\leavevmode
\includegraphics[height=15cm,angle=90,clip=true]{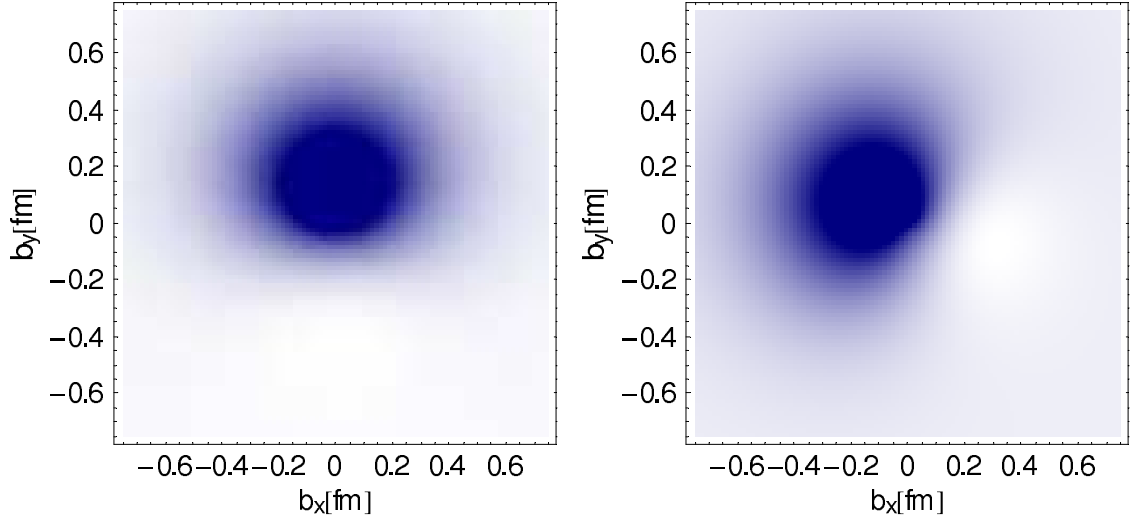}
\end{center}
\caption{\label{fig:impact-density-1} Illustration of the lowest
  moment $\half \int_{-1}^1 dx\, [F(x,\tvec{b}) + s^i
  F^i_T(x,\tvec{b})]$ for $u$-quarks in a proton with transverse spin
  vector $\tvec{S}=(1,0)$.  The transverse quark spin vector is
  $\tvec{s}=(1,0)$ in the left plot and $\tvec{s}=(0,1)$ in the right
  plot.}
\end{figure}

\begin{figure}
\begin{center}
\leavevmode
\includegraphics[height=15cm,angle=90,clip=true]{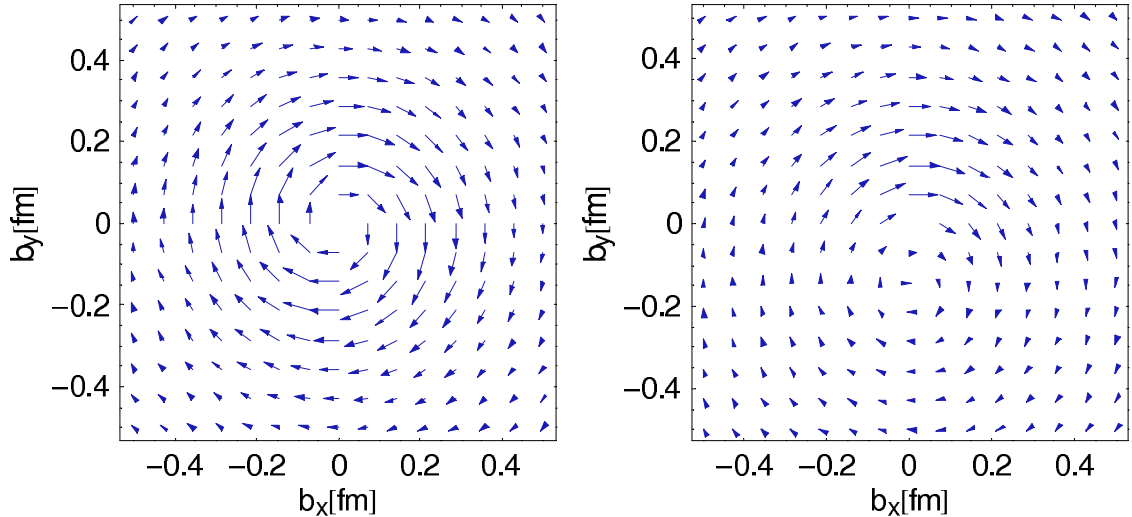}
\end{center}
\caption{\label{fig:impact-vectorfield-1} Illustration of the lowest
  moment $\int_{-1}^1 dx\, F^i_T(x,\tvec{b})$ of the vector field
  describing the transverse polarization of $u$-quarks in an
  unpolarized proton (left) and in a proton with transverse spin in
  the $x$-direction (right).}
\end{figure}

In Fig.~\ref{fig:impact-density-0} we show the resulting first moment
of the impact parameter density for unpolarized quarks in a
transversely polarized proton and for transversely polarized quarks in
an unpolarized proton.  The dipole-type structures due to $S^i
\epsilon^{ij} b^j E'$ and $s^i \epsilon^{ij} b^j (E_T' +
2\tilde{H}_T')$ are clearly visible, reflecting the large values in
(\ref{zero-values}) for $B_{10}$ and $B_{T 10} + 2\tilde{A}_{T 10}$ at
$t=0$.  The sum of both dipoles dominates the structure of the
distribution for transverse polarization of both quark and proton, as
is seen in Fig.~\ref{fig:impact-density-1}, whereas the
quadrupole-type term $s^i (2 b^i b^j - b^2 \delta^{ij}) S^j
\tilde{H}_T''$ is less prominent in our numerical example.  We note
that the two dipoles terms $S^i \epsilon^{ij} b^j E'$ and $s^i
\epsilon^{ij} b^j (E_T' + 2\tilde{H}_T')$ tend to cancel if quark and
proton spin are opposite to each other, and the resulting density (not
shown here) is rather sensitive to the precise values in the form
factor parameterizations.  Figure~\ref{fig:impact-vectorfield-1}
finally shows the lowest moment of the vector field
$F^i_T(x,\tvec{b})$ describing the transverse quark polarization in a
proton with or without transverse polarization.

\section{The spin matrix and positivity constraints}
\label{sec:positive}

\begin{figure}
\begin{center}
\leavevmode
\includegraphics[width=0.85\textwidth]{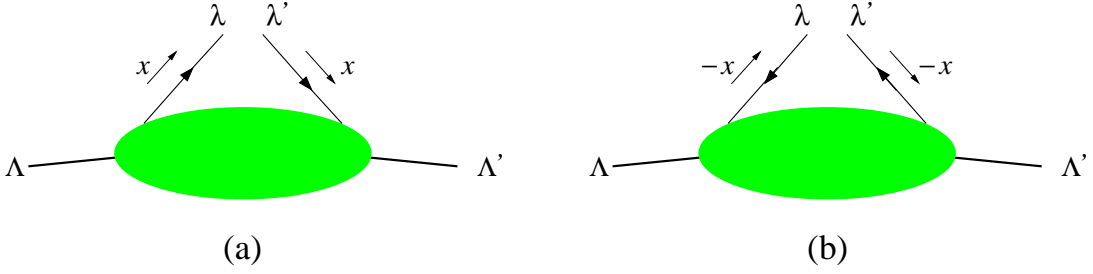}
\end{center}
\caption{\label{fig:handbag} Labeling of the proton and parton
  helicities in the matrix elements $M_{(\Lambda' \lambda') (\Lambda
  \lambda)}$ in the quark region $x>0$ (a) and the antiquark region
  $x<0$ (b).}
\end{figure}

In the previous section we have discussed the density of quarks with
transverse or longitudinal polarization in a transversely or
longitudinally polarized proton.  Densities for arbitrary polarization
states can be obtained from the spin matrix in the light-cone helicity
basis
\begin{equation}
  \label{spin-matrix-def}
M_{(\Lambda' \lambda') (\Lambda \lambda)}(x,\tvec{b}) =
\mathcal{N}^{-1}\, \int \frac{dz^-}{4\pi}\, e^{i x p^+ z^-}\,
  \Big\langle p^+,\tvec{0},\Lambda' \,\Big|\, 
  \bar{q}(z_2) \Gamma_{\lambda'\lambda}\,
  q(z_1) \Big|\, p^+,\tvec{0},\Lambda \Big\rangle
\end{equation}
with $z_1^+ =z_2^+ =0$, $\tvec{z}_1 =\tvec{z}_2 = \tvec{b}$, and
$z_1^- =-z_2^- =\half z^-$.  Here $\Lambda'$ and $\Lambda$ denote
light-cone helicities of the proton states, and definite light-cone
helicities $\lambda'$ and $\lambda$ of the quark are projected out by
the Dirac matrices (see e.g.\ \cite{Diehl:2001pm})
\begin{eqnarray}
  \label{quark-projectors}
\Gamma_{++} = \gamma^+ (1 + \gamma_5) , && \qquad\qquad
\Gamma_{-+} = - i\sigma^{+1} (1 + \gamma_5) , 
\nonumber \\
\Gamma_{--} = \gamma^+ (1 - \gamma_5) , && \qquad\qquad
\Gamma_{+-} = i\sigma^{+1} (1 - \gamma_5) .
\end{eqnarray}
The corresponding labeling of helicities in a handbag graph is shown
in Fig.~\ref{fig:handbag}a.  The matrix $M_{(\Lambda' \lambda')
(\Lambda \lambda)}$ reads
\begin{eqnarray}
  \label{spinm0}
\lefteqn{
\left( \renewcommand{\arraystretch}{2}
\begin{array}{cccc}
 H + \tilde H & -i e^{-i\phi} \displaystyle\frac{b}{m} E' &
 i e^{i\phi} \displaystyle\frac{b}{m} \Big(E_T'+2\tilde H_T'\Big) & 
 2\Big(H_T-\displaystyle\frac{1}{4m^2}\Delta_b \tilde H_T\Big)
\\
 i e^{i\phi} \displaystyle\frac{b}{m} E' & H - \tilde H & 
 2 e^{2 i\phi} \displaystyle\frac{b^2}{m^2} \tilde H_T'' &
 i e^{i\phi} \displaystyle\frac{b}{m} \Big(E_T'+2\tilde H_T'\Big)
\\
 -i e^{-i\phi} \displaystyle\frac{b}{m} \Big(E_T'+2\tilde H_T'\Big) &
 2 e^{-2i\phi} \displaystyle\frac{b^2}{m^2} \tilde H_T'' &
 H - \tilde H & -i e^{-i\phi} \displaystyle\frac{b}{m} E' 
\\
 2\Big(H_T-\displaystyle\frac{1}{4 m^2}\Delta_b \tilde H_T\Big) &
 -ie^{-i\phi} \displaystyle\frac{b}{m} \Big(E_T'+2\tilde H_T'\Big) &
 i e^{i\phi} \displaystyle\frac{b}{m} E' & H + \tilde H
\end{array}
\right)
}
\nonumber \\[0.5em]
&& \hspace{0.93\textwidth}
\end{eqnarray}
if we order the proton-quark helicity combinations as $(\Lambda
\lambda)= (++), (-+), (+-), (--)$.  Here GPDs are given in impact
parameter space with the notation specified after (\ref{trans-distr}),
and the azimuthal angle $\phi$ of $\tvec{b}$ is defined after
(\ref{density-terms}).  The quark density for an arbitrary
polarization state of proton and quark can be written as $\half
\sum_{\Lambda'\lambda'\Lambda\lambda}\,
(c_{\Lambda'\lambda'\phantom{\!\!\!\!\!()}})^* M_{(\Lambda' \lambda')
(\Lambda \lambda)}\, c_{\Lambda\lambda\phantom{\!\!\!\!\!()}}$ with
coefficients normalized to $\sum_{\Lambda\lambda} | c_{\Lambda\lambda}
|^2 = 1$.  This implies that the matrix $M_{(\Lambda' \lambda')
(\Lambda \lambda)}$ must be positive semidefinite.  Integration of
(\ref{spinm0}) over $\tvec{b}$ leads to the known spin matrix for the
forward distribution functions $f_1(x)$, $g_1(x)$, $h_1(x)$ according
to (\ref{forward-limits}), and the positivity of the corresponding
eigenvalues gives immediately the Soffer bound $2 |h_1(x)| \le
f_1(x)+g_1(x)$.  Using the relations (\ref{dictionary}) we see that
the matrix (\ref{spinm0}) of impact parameter dependent distributions
is the exact analog of the spin matrix for transverse momentum
dependent distribution functions, which was discussed in
\cite{Bacchetta:1999kz}.\footnote{To compare with the matrices given
in \protect\cite{Bacchetta:1999kz} one must take into account that in
those papers the sign convention for the Sivers function
$f_{1T}^\perp$ is opposite to the convention from
\protect\cite{Boer:1997nt} used here, and that the rows of the
matrices in those papers correspond to the indices $(\Lambda \lambda)$
rather than $(\Lambda' \lambda')$.  We thank A.~Bacchetta for
clarifying discussions on this issue.}

In order to simplify the following discussion, we change basis by
multiplying (\ref{spinm0}) with the diagonal matrix $D =
\mathrm{diag}(1, i e^{i\phi}, -i e^{-i\phi}, 1)$ from the right and
with $D^\dag$ from the left.  This gives a matrix
\begin{equation}
  \label{spinm1}
\left( \renewcommand{\arraystretch}{2}
\begin{array}{cccc} 
 H + \tilde H & \displaystyle\frac{b}{m} E' &
 \displaystyle\frac{b}{m} \Big(E_T'+2\tilde H_T'\Big) & 
 2\Big(H_T-\displaystyle\frac{1}{4m^2}\Delta_b \tilde H_T\Big)
\\
 \displaystyle\frac{b}{m} E' & H - \tilde H &
 -2 \displaystyle\frac{b^2}{m^2} \tilde H_T'' &
 \displaystyle\frac{b}{m} \Big(E_T'+2\tilde H_T'\Big)
\\
 \displaystyle\frac{b}{m} \Big(E_T'+2\tilde H_T'\Big) &
 -2 \displaystyle\frac{b^2}{m^2} \tilde H_T'' &
 H - \tilde H & \displaystyle\frac{b}{m} E' 
\\
 2\Big(H_T-\displaystyle\frac{1}{4 m^2}\Delta_b \tilde H_T\Big) &
 \displaystyle\frac{b}{m} \Big(E_T'+2\tilde H_T'\Big) &
 \displaystyle\frac{b}{m} E' & H + \tilde H   \; ,
\end{array}
\right)
\end{equation}
which is purely real and depends on $b$ but no longer on $\phi$.
Positivity of the upper left $2\times 2$ sub-matrix of (\ref{spinm1})
leads to the bound
\begin{equation}
\frac{b}{m}\, |E'| \le \sqrt{H^2-\tilde H^2} ,
\label{bound1}
\end{equation}
which has been discussed in \cite{Burkardt:2003ck}.  Using the
eigenvalues of the full matrix (\ref{spinm1}), we can tighten these
bounds by including the tensor GPDs. With the abbreviations
\begin{eqnarray}
a_\pm&=&H \pm \Big( H_T + \frac{1}{m^2} \tilde H_T'
          - \frac{1}{2m^2}\Delta_b \tilde H_T \Big) \;,
\nonumber\\
b_\pm&=& \tilde H \pm \Big( H_T - \frac{1}{m^2}\tilde H_T' \Big) \;,
\nonumber\\
c_\pm&=& \frac{b}{m} \Big(E' \pm E_T' \pm  2\tilde H_T'\Big)
\label{short1}
\end{eqnarray}
the four eigenvalues read
\begin{eqnarray}
a_+ + \sqrt{b_+^2+c_+^2} \, , \qquad
a_+ - \sqrt{b_+^2+c_+^2} \, , \qquad
a_- + \sqrt{b_-^2+c_-^2} \, , \qquad
a_- - \sqrt{b_-^2+c_-^2} \, .
\label{EVs1}
\end{eqnarray}
We see that they are related pairwise by changing the sign of all
chiral-odd distributions.  This is tantamount to multiplying the
matrix (\ref{spinm1}) with $\mathrm{diag}(1,1,-1,-1)$ from the left
and from the right, which does of course not change its eigenvalues.
Positivity of the eigenvalues (\ref{EVs1}) gives the bounds
\begin{equation}                                                    
0 \:\le\: a_\pm \, , \qquad\qquad
c_\pm^2 \:\le\: a_\pm^2 - b_\pm^2 
        \:=\: (a_\pm+b_\pm)(a_\pm-b_\pm) \, ,
\label{bound}
\end{equation}
which in particular imply $0 \le a_\pm+b_\pm$ and $0 \le a_\pm-b_\pm$.
They explicitely read
\begin{equation}
  \label{bound2}
\Big|\, H_T + \frac{1}{m^2} \tilde H_T' 
        - \frac{1}{2m^2}\Delta_b \tilde H_T \,\Big| \le H
\end{equation}
and
\begin{eqnarray}
  \label{bound2a}
\frac{b^2}{m^2} \Big( E' \pm E_T' \pm  2\tilde H_T' \Big)^2 &\le &
\Big( H \pm H_T \pm \frac{1}{m^2}\tilde H_T' 
   \mp \frac{1}{2m^2}\Delta_b \tilde H_T \Big)^2
 - \Big( \tilde H \pm H_T \mp \frac{1}{m^2} \tilde H_T' \Big)^2 \, .
\hspace{1em}
\end{eqnarray}
In the phenomenological study \cite{Diehl:2004cx} it was found that
the bound (\ref{bound1}) can indeed be very restrictive (and thus
helpful) in reconstructing generalized parton distributions from
experimental data.  The tighter bounds (\ref{bound2}) and
(\ref{bound2a}) may therefore be of practical value even with limited
information on the three additional chiral-odd distributions they
contain.  As in the case of the usual parton distributions,
renormalization of the operators in (\ref{spin-matrix-def}) may
destroy the density interpretation of the impact parameter
distributions.  Closer analysis reveals that the bounds following from
positivity of the matrix $M_{(\Lambda' \lambda') (\Lambda \lambda)}$
should be valid at a sufficiently high renormalization scale $\mu^2$
\cite{Pobylitsa:2002ru}.  They are stable under leading order
evolution to higher scales, as shown in \cite{Pobylitsa:2002iu}.

Since both experimental information and results from lattice QCD
calculations are in the first instance given as a function of $t$, it
is useful to have bounds also directly in momentum space.  This can be
achieved by applying to (\ref{bound2a}) the method used in
\cite{Burkardt:2003ck} for the simpler bound (\ref{bound1}), where the
main ingredient is the Schwarz inequality.  A method leading to the
same results is to multiply (\ref{spinm1}) from the left and the right
with $\mathrm{diag}(1, mb,mb, 1)$.  In the resulting matrix
$\widehat{M}$ only even powers of $b$ appear.  Integrating over $b$ as
$2\pi \int_0^\infty db\, b\, \widehat{M}(b^2) = \int d^2\tvec{b}\,
\widehat{M}(b^2) $, one obtains GPDs and their derivatives at zero
momentum transfer $t=0$.  The result is still a positive semidefinite
matrix, whose eigenvalues have the form (\ref{EVs1}) with
\begin{eqnarray}
a_\pm &=& \Big[\, H + \tilde H
          + 4m^2 \frac{\partial}{\partial t} (H - \tilde H)
          \pm 2 (H_T - 2\tilde H_T) \,\Big]_{t=0} \; ,
\nonumber\\
b_\pm &=& \Big[\, H + \tilde H
          - 4m^2 \frac{\partial}{\partial t} (H - \tilde H)
          \pm 2 (H_T + 2\tilde H_T) \,\Big]_{t=0} \; ,
\nonumber\\
c_\pm &=& \Big[\, 2 (E \pm E_T \pm 2 \tilde H_T) \,\Big]_{t=0} \; .
\phantom{\frac{\partial}{\partial t}}
\label{short2}
\end{eqnarray}
This leads to the bounds
\begin{equation}
  \label{bound3a}
\Bigg[ \Big( E \pm E_T \pm 2 \tilde H_T \Big)^2 \,\Bigg]_{t=0} \le
\Bigg[ \Big( H + \tilde H \pm 2 H_T \Big)\,
\Big( 4m^2 \frac{\partial}{\partial t} (H - \tilde H)
      \mp 4\tilde H_T \Big) \,\Bigg]_{t=0} \, ,
\end{equation}
where both expressions in large parentheses on the right-hand side
must be positive or zero according to our remark after (\ref{bound}).
The condition $0 \le [H + \tilde H \pm 2 H_T]_{t=0}$ is just the
Soffer bound.

Alternatively, we can multiply the matrix (\ref{spinm1}) from the left
and the right with $\mathrm{diag}(mb, 1,1, mb)$ and then integrate
over $b$ as described above.  The eigenvalues have again the structure
of (\ref{EVs1}), and we obtain bounds
\begin{equation}
  \label{bound4a}
\Bigg[ \Big( E \pm E_T \pm 2 \tilde H_T \Big)^2 \,\Bigg]_{t=0} \le
\Bigg[ \Big( H - \tilde H
      \mp \frac{1}{2m^2} \int_{-\infty}^t dt'\, \tilde H_T(t') \Big)\,
      \Big( 4m^2\frac{\partial}{\partial t} (H +\tilde H \pm 2H_T)
      \mp 2\tilde H_T \Big) \,\Bigg]_{t=0}
\end{equation}
with both expressions in large parentheses on the right-hand side
positive or zero.  They contain a function which is non-local in
momentum space, namely
\begin{equation}
\int_{-\infty}^0 dt \tilde{H}_T(t) = 4\pi \tilde{H}_T(b=0) ,
\end{equation}
which can be traced back to the integral $\int d^2\tvec{b}\; b^2
\tilde{H}_T''(b^2)$ in the derivation.

For reasons which will become clear shortly, certain applications
require bounds which do not involve the distribution $\tilde{H}$ .
One such bound is simply obtained by omitting the last term in
(\ref{bound2a}), which together with (\ref{bound2}) leads to
\begin{eqnarray}
  \label{bound5}
\frac{b}{m}\, \Big| E' \pm E_T' \pm  2\tilde H_T' \Big| &\le &
H \pm \Big( H_T + \frac{1}{m^2}\tilde H_T' 
  - \frac{1}{2m^2}\Delta_b \tilde H_T \Big) \, .
\end{eqnarray}
To obtain a bound in momentum space we multiply (\ref{bound5}) with $m
b$, integrate over $\tvec{b}$, and use the Schwarz inequality in the
forms $\int d^2\tvec{b}\, f \le \int d^2\tvec{b}\, |f|$ and $\int
d^2\tvec{b}\, b\, g \le (\int d^2\tvec{b}\, g)^{1/2}\, (\int
d^2\tvec{b}\, b^2 g)^{1/2}$, following the method of
\cite{Burkardt:2003ck}.  This leads to
\begin{equation}
  \label{bound5a}
\Bigg[ \Big( E \pm E_T \pm 2 \tilde H_T \Big)^2 \,\Bigg]_{t=0} \le
\Bigg[ \Big( H \pm H_T 
  \mp \frac{1}{4m^2} \int_{-\infty}^t dt'\, \tilde H_T(t') \Big)\,
  \Big( 4m^2\frac{\partial}{\partial t} (H \pm H_T)
      \mp 3\tilde{H}_T \Big) \,\Bigg]_{t=0} \; , \hspace{1em}
\end{equation}
where both terms in large parentheses on the right-hand side must be
positive or zero.  Alternatively one can add to (\ref{spinm1}) the
positive semidefinite matrix
\begin{equation}
  \label{spinm2}
\left( \renewcommand{\arraystretch}{2}
\begin{array}{cccc} 
 H - \tilde H & \displaystyle\frac{b}{m} E' & 0 & 0 \\
 \displaystyle\frac{b}{m} E' & H + \tilde H & 0 & 0 \\
 0 & 0 & H + \tilde H & \displaystyle\frac{b}{m} E' \\
 0 & 0 & \displaystyle\frac{b}{m} E' & H - \tilde H
\end{array}
\right)
\end{equation}
and proceed as above.  One then obtains bounds analogous to
(\ref{bound2a}) and to (\ref{bound3a}) and (\ref{bound4a}) by the
replacements $H \to 2 H$, $E \to 2 E$, $\tilde{H} \to 0$.  They
explicitly read
\begin{eqnarray}
  \label{bound2b}
\frac{b^2}{m^2} \Big( 2 E' \pm E_T' \pm  2\tilde H_T' \Big)^2 &\le &
\Big( 2 H \pm H_T \pm \frac{1}{m^2}\tilde H_T' 
   \mp \frac{1}{2m^2}\Delta_b \tilde H_T \Big)^2
 - \Big( H_T - \frac{1}{m^2} \tilde H_T' \Big)^2
\end{eqnarray}
and
\begin{eqnarray}
  \label{bound3b}
\Bigg[ \Big( 2 E \pm E_T \pm 2 \tilde H_T \Big)^2 \,\Bigg]_{t=0} 
&\le &
\Bigg[ 4\, \Big( H \pm H_T \Big)\,
      \Big( 4m^2 \frac{\partial}{\partial t} H
      \mp 2\tilde H_T \Big) \,\Bigg]_{t=0} \, ,
\nonumber \\
\Bigg[ \Big( 2 E \pm E_T \pm 2 \tilde H_T \Big)^2 \,\Bigg]_{t=0} 
&\le &
\Bigg[ 4\, \Big( H
      \mp \frac{1}{4m^2} \int_{-\infty}^t dt'\, \tilde H_T(t') \Big)\,
\Big( 4m^2\frac{\partial}{\partial t} (H \pm H_T)
      \mp \tilde H_T \Big) \,\Bigg]_{t=0} \, . \hspace{1em}
\hspace{2em}
\end{eqnarray}

So far we have considered quark distributions.  For antiquarks we
define the matrix $M_{(\Lambda' \lambda') (\Lambda
\lambda)}(x,\tvec{b})$ with $x<0$ as in (\ref{spin-matrix-def}), but
with the Dirac matrices $\Gamma_{\lambda'\lambda}$ now reading
\begin{eqnarray}
  \label{antiquark-projectors}
\Gamma_{++} = - \gamma^+ (1 - \gamma_5) , && \qquad\qquad
\Gamma_{-+} = i\sigma^{+1} (1 + \gamma_5) , 
\nonumber \\
\Gamma_{--} = - \gamma^+ (1 + \gamma_5) , && \qquad\qquad
\Gamma_{+-} = - i\sigma^{+1} (1 - \gamma_5)
\end{eqnarray}
instead of (\ref{quark-projectors}).  A global minus sign compared
with the quark case arises because the order of the operators
$\bar{q}$ and $q$ in (\ref{spin-matrix-def}) has to be reversed to
obtain a density operator for antiquarks.  To understand the signs in
front of $\gamma_5$, recall that antiquarks with positive helicity
have negative chirality.  One must finally keep in mind that the
helicity index $\lambda$ refers to the parton on the left-hand side of
the handbag diagram as shown in Fig.~\ref{fig:handbag}.  For
antiquarks this parton is annihilated by the operator $\bar{q}$, and
not by $q$ as for quarks.  Comparing (\ref{antiquark-projectors}) with
(\ref{quark-projectors}), we find that the spin matrix $M_{(\Lambda'
\lambda') (\Lambda \lambda)}(x,\tvec{b})$ in the antiquark region
$x<0$ reads as in (\ref{spinm0}), but with the signs of all GPDs
except $\tilde{H}$ reversed.  It is positive semidefinite, and one
readily obtains bounds for antiquarks analogous to those we have given
for quarks.

Let us finally address the question of positivity bounds for Mellin
moments of generalized parton distributions at $\xi=0$, which are for
instance relevant in lattice QCD calculations.  As discussed in the
previous section, these moments involve both the quark and antiquark
regions, $x>0$ and $x<0$.  Clearly, it is only the sum of quark and
antiquark densities and not their difference for which positivity is
ensured.  This leads us to consider the moments $\int_{-1}^1 dx\,
x^{n-1} f(x,b^2)$ with even $n$ for all distributions $f$ except
$\tilde{H}$, which is why we have derived bounds without $\tilde{H}$.
To derive bounds for the even moments, we can add the positive
semidefinite matrices $\int_{0}^1 dx\, x^{n-1} M(x,\tvec{b})$ and
$\int_{-1}^0 dx\, (-x)^{n-1} M(x,\tvec{b})$.  The result involves
Mellin moments $\int_{-1}^1 dx\, x^{n-1} f(x,b^2)$ for all
distributions $f$ except $\tilde{H}$, where instead one has
$\int_{-1}^1 dx\, x^{n-1}\, \mbox{sgn}(x) \tilde{H}(x,b^2)$, which is
the matrix element of a highly nonlocal operator.  Positivity bounds
are then obtained exactly as above, and we find that the inequalities
(\ref{bound5}), (\ref{bound5a}) and (\ref{bound2b}), (\ref{bound3b})
also hold when we replace all distributions with their even Mellin
moments.

It may be interesting to see whether the bounds which do involve
$\tilde{H}$ also hold when we replace this distribution by its Mellin
moment $\int_{-1}^1 dx\, x^{n-1} \tilde{H}(x,b^2)$, thus taking the
``wrong'' sign in the antiquark region $x<0$, or whether the bounds
given in this section also hold for odd Mellin moments.  This would
signal that the antiquark contribution to the moments in question is
sufficiently small to not destroy positivity of the quark
contribution, i.e.\ of the matrix $\int_{0}^1 dx\, x^{n-1}
M(x,\tvec{b})$.


\section{Equations of motion and distributions of twist three}
\label{sec:eom}

At the end of Sect.~\ref{sec:transv} we have encountered the total
derivatives of the chiral-odd operators which define transversity
distributions through the matrix element $F_T^j$.  Using the Dirac
equation for the quark field operator, we can rewrite the local
operators appearing in (\ref{mellin-derivative}) as
\begin{eqnarray}
  \label{eom-g5}
\partial_\mu \Big[
  \bar{q} \sigma^{+\mu} \gamma_5 (i\lrD^+)^{n-1} q \Big]
 &=& -2\, \bar{q} (i\lrD^+)^{n} \gamma_5 q
     + \sum_{i=2}^{n} \bar{q} (i\lrD^+)^{i-2}\,
       \sigma^{+}{}_{\mu} \gamma_5\, g G^{\mu +} (i\lrD^+)^{n-i} q \, ,
\\
  \label{eom-1}
\partial_\mu \Big[
  \bar{q} \sigma^{+\mu} (i\lrD^+)^{n-1} q \Big]
 &=& -2\, \bar{q} (i\lrD^+)^{n} q
     + \sum_{i=2}^{n} \bar{q} (i\lrD^+)^{i-2}\,
       \sigma^{+}{}_{\mu}\, g G^{\mu +} (i\lrD^+)^{n-i} q
\nonumber \\
 && {}+ 2 m_q\, \bar{q} \gamma^+ (i\lrD^+)^{n-1} q 
\hspace{2em}
\end{eqnarray} 
for $n\ge 1$, where we have used $\partial_\mu [\bar{q} \sigma^{+\mu}
\ldots q] = -i \bar{q}\, ( \fslash{\lD}\, \gamma^+ + 2\lrD^+ -
\gamma^+\!  \fslash{\rD} \,) \ldots q$ and $[\rD^\mu, \lrD^+] = -i g
G^{\mu+}$.  Apart from the term proportional to the quark mass $m_q$,
the operators on the right-hand sides are of twist three.  They are
obtained by inserting covariant derivatives $i \lrD^+$ into the
pseudoscalar or scalar quark current and into the
quark-antiquark-gluon operators $\bar{q} \sigma^{+}{}_{\mu} \gamma_5\,
g G^{\mu +} q$ or $\bar{q} \sigma^{+}{}_{\mu}\, g G^{\mu +} q$ (which
can be written in a number of ways using the relations
$\sigma^{\lambda\mu} \gamma_5 = -\smash{\half} i
\epsilon^{\lambda\mu\alpha\beta} \sigma_{\alpha\beta}$ and
$\tilde{G}^{\lambda\mu} = \smash{\half}
\epsilon^{\lambda\mu\alpha\beta} G_{\alpha\beta}$).  These
quark-antiquark-gluon operators are chiral-odd partners of the
operators obtained by inserting covariant derivatives into $\bar{q}
\gamma^+ g G^{\mu +} q$ and $\bar{q} \gamma^+ \gamma_5\, g
\tilde{G}^{\mu +} q$, which appear in the virtual Compton amplitude at
twist-three accuracy and are well-known from inclusive deep inelastic
scattering and from deeply virtual Compton scattering, see e.g.\
\cite{Belitsky:2000vx}.  The forward matrix elements of the operators
in (\ref{eom-1}) appear for instance in Drell-Yan pair production and
have been studied in detail in \cite{Jaffe:1991ra}.  A review of their
properties, including their renormalization group evolution, can be
found in~\cite{Kodaira:1998jn}.  Note that the derivative operator on
the left-hand side of (\ref{eom-1}) does not contribute to forward
matrix elements.

The powers of covariant derivatives in (\ref{eom-g5}) and
(\ref{eom-1}) can be resummed to obtain nonlocal operators on the
light-cone, which may be written as
\begin{eqnarray}
\mathcal{O}_{2}(x) &=& 
  \int_{-\infty}^{\infty} \frac{dz^-}{2\pi}\, e^{ix P^+ z^-}\;
  \bar{q}(-\half z) W[-\half z, \half z]\,
  \Gamma q(\half z)
\nonumber \\
\mathcal{O}_{3}(x) &=& -i
  \int_{-\infty}^{\infty} \frac{dz^-}{2\pi}\, e^{ix P^+ z^-}
  \int_{-\frac{1}{2} z^-}^{\frac{1}{2} z^-} dy^- \;
  \bar{q}(-\half z) W[-\half z, y]\,
  \Gamma_\mu\, gG^{\mu +}(y)\, W[y, \half z]\,
  q(\half z) ,
\hspace{2em}
\end{eqnarray}
where $z^+ = y^+ = 0$ and $\tvec{z} = \tvec{y} = 0$, the Wilson lines
$W$ are along light-like paths, and $\Gamma$ and $\Gamma_\mu$ denote
Dirac matrices.  Local operators at position $z=0$ are then obtained
by
\begin{eqnarray}
(P^+)^n \int dx\, x^{n-1}\, \mathcal{O}_{2}(x) &=& 
  \bar{q}\, (i \lrD^+)^{n-1} \Gamma q
\nonumber \\
(P^+)^n \int dx\, x^{n-1}\, \mathcal{O}_{3}(x) &=& 
\sum_{i=2}^{n} \bar{q}\, (i \lrD^+)^{i-2}\,
  \Gamma_\mu\, gG^{\mu +} (i \lrD^+)^{n-i} q
\end{eqnarray}
for $n\ge 1$.  Note that the integral $\int dx\, \mathcal{O}_{3}(x)$
is zero.  The matrix elements of the nonlocal operators
$\mathcal{O}_{2}(x)$ and $\mathcal{O}_{3}(x)$ between nucleon states
are parameterized by suitable generalized parton distributions.
Taking instead matrix elements between the vacuum and a meson state
one obtains meson distribution amplitudes, and nonlocal versions of
the equations of motion in (\ref{eom-g5}) and (\ref{eom-1}) have been
extensively used in this context \cite{Ball:1998sk,Ball:1998je}.  We
note that the operators $\mathcal{O}_{3}(x)$ with $\Gamma^\mu=
\sigma^{+\mu}$ or $\sigma^{+\mu} \gamma_5$ involve only ``good''
components of the quark and gluon fields in the parlance of light-cone
quantization and hence admit an interpretation in terms of parton
degrees of freedom, unlike the operators $\mathcal{O}_{2}(x)$ with
$\Gamma =1$ or $\gamma_5$, which are products of one ``good'' and one
``bad'' field component \cite{Jaffe:1991ra,Jaffe:1996zw}.

With possible applications to lattice QCD calculations in mind, we
prefer here to work with the local operators in (\ref{eom-g5}) and
(\ref{eom-1}) and the form factors parameterizing the Mellin moments
of GPDs.  Instead of transforming (\ref{mellin-derivative}) back from
impact parameter to momentum space, we can directly use translation
invariance to obtain $\langle p'| \partial_\mu \mathcal{O} | p\rangle
= i \Delta_\mu \langle p'| \mathcal{O} | p\rangle$ for a local
operator $\mathcal{O}$.  {}From (\ref{eom-g5}) and (\ref{eom-1}) we
then obtain relations between the form factors of twist-two and
twist-three operators.  We give results for $n=1$ and $n=2$, their
generalization to higher moments is straightforward.  In contrast to
the previous sections, we consider the general case where $\xi$ need
not be zero.  Using the constraints from parity and time reversal
invariance, the quark tensor current can be parameterized by
\begin{eqnarray}
  \langle p'| \bar{q} \sigma^{\lambda\mu} \gamma_5\, q | p\rangle
 &=& \bar{u} \sigma^{\lambda\mu} \gamma_5\, u \, A_{T 10}(t)
  + \bar{u} \frac{\epsilon^{\lambda\mu\alpha\beta}
    \Delta_\alpha P_\beta}{m^2} u\, \tilde{A}_{T 10}(t)
  + \bar{u} \frac{\epsilon^{\lambda\mu\alpha\beta}
    \Delta_\alpha \gamma_\beta}{2m} u\, B_{T 10}(t) ,
\end{eqnarray}
where we use the notation of \cite{Hagler:2004yt}.  An analog for the
operator $\bar{q} \sigma^{\lambda\mu} q$ is readily obtained using
$\sigma^{\lambda\mu} \gamma_5 = -\half i
\epsilon^{\lambda\mu\alpha\beta} \sigma_{\alpha\beta}$.  For the
operator with one covariant derivative we have
\begin{eqnarray}
  \label{tensor-ffs}
\mathbf{A}_{\lambda\mu} \mathbf{S}_{\mu\nu}
  \langle p'| \bar{q} \sigma^{\lambda\mu} \gamma_5\, 
      i\lrD^\nu q | p\rangle
 &=& \mathbf{A}_{\lambda\mu} \mathbf{S}_{\mu\nu} \Big[
    \bar{u} \sigma^{\lambda\mu} \gamma_5\, u\, P^\nu A_{T 20}(t)\, 
  + \bar{u} \frac{\epsilon^{\lambda\mu\alpha\beta}
    \Delta_\alpha P_\beta}{m^2} u\, P^\nu \tilde{A}_{T 20}(t)
\nonumber \\[0.2em]
 && {}+ \bar{u} \frac{\epsilon^{\lambda\mu\alpha\beta}
    \Delta_\alpha \gamma_\beta}{2m} u\, P^\nu B_{T 20}(t)
  + \bar{u} \frac{\epsilon^{\lambda\mu\alpha\beta}
    P_\alpha \gamma_\beta}{m} u\, \Delta^\nu \tilde{B}_{T 21}(t) 
 \Big] ,
\end{eqnarray}
where $\mathbf{A}_{\lambda\mu}$ denotes antisymmetrization in
$\lambda$ and $\mu$ and $\mathbf{S}_{\mu\nu}$ denotes symmetrization
and subtraction of the trace.  Comparison with (\ref{gpd-def}) readily
gives
\begin{eqnarray}
\int_{-1}^1 dx\, x^{n-1} H_T(x,\xi,t) = A_{T n0}(t) ,
&& \qquad
\int_{-1}^1 dx\, x^{n-1} \tilde{H}_T(x,\xi,t) = \tilde{A}_{T n0}(t) ,
\nonumber \\
\int_{-1}^1 dx\, x^{n-1} E_T(x,\xi,t) = B_{T n0}(t) ,
&& \qquad
\int_{-1}^1 dx\, x \tilde{E}_T(x,\xi,t) = -2\xi \tilde{B}_{T 21}(t)
\end{eqnarray}
for $n=1,2$.  In the forward limit we have $H_T(x,0,0) = h_1(x)$, so
that $A_{T n0}(0) = \int_{-1}^1 dx\, x^{n-1} h_1(x)$ is a moment of
the usual transversity distribution.  The contractions
\begin{equation}
\Delta_\mu\langle p'|
  \bar{q} \sigma^{+\mu} \gamma_5 (i\lrD^+)^{n-1} q | p\rangle \, ,
\qquad\qquad\qquad
\Delta_\mu \langle p'| 
  \bar{q} \sigma^{+\mu} (i\lrD^+)^{n-1} q | p \rangle
\end{equation}
needed for the equation of motion constraints (\ref{eom-g5}) and
(\ref{eom-1}) respectively project out the form factors  $A_{T ni}$,
$\tilde{B}_{T ni}$ and $A_{T ni}$, $\tilde{A}_{T ni}$, $B_{T ni}$.

For the twist-two operators constructed from the quark vector
current we have
\begin{eqnarray}
  \langle p'| \bar{q} \gamma^\mu\, q | p\rangle
 &=& \bar{u} \gamma^\mu u\, A_{10}(t)
   + \bar{u} \frac{i \sigma^{\mu\alpha} 
        \Delta_\alpha}{2m} u\, B_{10}(t) \, ,
\nonumber \\
\mathbf{S}_{\mu\nu} \langle p'| \bar{q} \gamma^\mu\,
      i\lrD^\nu q | p\rangle
 &=& \mathbf{S}_{\mu\nu}  \Big[
     \bar{u} \gamma^\mu u\, P^\nu A_{20}(t)
   + \bar{u} \frac{i \sigma^{\mu\alpha} 
        \Delta_\alpha}{2m} u\, P^\nu B_{20}(t)
   + \bar{u} \frac{\Delta^\mu \Delta^\nu}{m} u\, C_2(t) \Big]
\end{eqnarray}
and
\begin{eqnarray}
\int_{-1}^1 dx\, H(x,\xi,t) &=& A_{10}(t) ,
\qquad\qquad\qquad \hspace{1em}
\int_{-1}^1 dx\, E(x,\xi,t) \;=\; B_{10}(t) ,
\nonumber \\
\int_{-1}^1 dx\, x H(x,\xi,t) &=& A_{20}(t) + 4\xi^2 C_2(t),
\qquad
\int_{-1}^1 dx\, x E(x,\xi,t) \;=\; B_{20}(t) - 4\xi^2 C_2(t) .
\end{eqnarray}
Note that $A_{10}(t)$ and $B_{10}(t)$ simply are the contributions of
the relevant quark flavor to the usual Dirac and Pauli form factors,
respectively.  In the forward limit $A_{n0}(0) = \int_{-1}^1 dx\,
x^{n-1} f_1(x)$ is a moment of the unpolarized parton distribution.

We further parameterize the twist-three operators constructed from the
quark scalar and pseudoscalar currents as
\begin{eqnarray}
  \label{scalar-ffs}
  \langle p'| \bar{q}\, i\lrD^\mu q | p\rangle
 &=& m\, \bar{u} \gamma^\mu u\, A_{S 10} 
  + \half \bar{u} i\sigma^{\mu\alpha} \Delta_\alpha u\, B_{S 10} \, ,
\nonumber \\
\mathbf{S}_{\mu\nu} 
  \langle p'| \bar{q}\, i\lrD^\mu\, i\lrD^\nu q | p\rangle
 &=& \mathbf{S}_{\mu\nu} \Big[
    m\, \bar{u} \gamma^\mu u\, P^\nu\, A_{S 20} 
  + \half \bar{u} i\sigma^{\mu\alpha} \Delta_\alpha u\, 
    P^\nu\, B_{S 20}
  + \bar{u} u\, \Delta^\mu \Delta^\nu\, C_{S 2} \,\Big] ,
\nonumber \\
  \langle p'| \bar{q}\, i\lrD^\mu \gamma_5 q | p\rangle
 &=& \half \bar{u} \gamma_5 u\, P^\mu\, \tilde{B}_{P 10} \, ,
\nonumber \\
\mathbf{S}_{\mu\nu}
  \langle p'| \bar{q}\, i\lrD^\mu\, i\lrD^\nu \gamma_5 q | p\rangle
 &=& \mathbf{S}_{\mu\nu} \Big[ 
    m\, \bar{u} \gamma^\mu \gamma_5 u\, \Delta^\nu\, \tilde{A}_{P 21} 
  + \half \bar{u} \gamma_5 u
    \Big( P^\mu P^\nu \tilde{B}_{P 20} 
        + \Delta^\mu \Delta^\nu\tilde{B}_{P 22} \Big) \Big] ,
\hspace{1.5em}
\end{eqnarray} 
where we have omitted the argument $t$ of the form factors for
brevity.  Following \cite{Ji:1998pc,Hagler:2004yt} we have assigned
the subscripts of form factors such that the first subscript gives the
spin of the operator (i.e.\ the number of Lorentz indices in the
symmetrization and subtraction of traces).  The second subscript
counts the number of factors $\Delta$ in the form factor decomposition
whose Lorentz index corresponds to a covariant derivative on the
operator side.  In the forward limit $p=p'$ only the form factors
$A_{S n0}(t)$ survive in (\ref{scalar-ffs}).  They are the moments of
the chiral-odd parton distribution $e(x)$ defined in
\cite{Jaffe:1991ra,Mulders:1995dh}, given by $A_{S n0}(0) =
\int_{-1}^1 dx\, x^{n} e(x)$.  Note that the local current $\bar{q} q$
without a covariant derivative (whose forward matrix element is
related to the pion-nucleon sigma-term) does not appear in the
constraints (\ref{eom-1}).  In other words, the equation of motion
constraint involves $x e(x)$ when resummed to $x$ space, and thus is
not affected by the $\delta(x)$ singularity of $e(x)$, discussed e.g.\
in the recent review~\cite{Efremov:2002qh}.

If we finally define form factors for the quark-antiquark-gluon matrix
elements as
\begin{eqnarray}
\mathbf{S}_{\mu\nu} \langle p'| \bar{q}
   \sigma^{\mu}{}_\alpha\, g G^{\alpha\nu} q | p\rangle
 &=& \mathbf{S}_{\mu\nu} \Big[
     2m\, \bar{u} \gamma^\mu u\, P^\nu\, A_{G 20} 
   + \bar{u} i\sigma^{\mu\alpha} \Delta_\alpha u\, P^\nu \, B_{G 20}
   + 2 \bar{u} u\, \Delta^\mu \Delta^\nu\, C_{G 2} \Big] ,
\nonumber \\
\mathbf{S}_{\mu\nu} \langle p'| \bar{q}
   \sigma^{\mu}{}_\alpha \gamma_5\, g G^{\alpha\nu} q | p\rangle
 &=& \mathbf{S}_{\mu\nu} \Big[
    2m\, \bar{u} \gamma^\mu \gamma_5 u\, \Delta^\nu\, \tilde{A}_{G 21}
  + \bar{u}\gamma_5 u\, 
    \Big( P^\mu P^\nu\, \tilde{B}_{G 20}
        + \Delta^\mu \Delta^\nu\, \tilde{B}_{G 22} \Big) \Big] ,
\hspace{2em}
\end{eqnarray}
the equations of motion embodied in (\ref{eom-g5}) and (\ref{eom-1})
give relations
\begin{eqnarray}
  \label{first-moments}
\rule{0pt}{2em}
A_{S 10} &=& \frac{m_q}{m}\, A_{10}
  - \frac{t}{4m^2}\, \Big( B_{T 10} + 2 \tilde{A}_{T 10} \Big) ,
\nonumber \\
B_{S 10} &=& \frac{m_q}{m}\, B_{10}
  - \Big( A_{T 10} - \frac{t}{2m^2} \tilde{A}_{T 10} \Big) ,
\nonumber \\
\tilde{B}_{P 10} &=& 2 A_{T 10}   \phantom{\frac{t}{2m^2}}
\rule[-1em]{0pt}{1em}
\end{eqnarray}
for the lowest moments.  At this level the quark-antiquark-gluon
operators do not appear yet.  In particular, at $t=0$ the first
relation in (\ref{first-moments}) gives the well-known sum rule
$\int_{-1}^1 dx\, x e(x) = (m_q/m)\, \smash{\int_{-1}^{1}} dx\,
f_1(x)$, where the integral over $f_1(x)$ at the right-hand side is
just the number of valence quarks with appropriate flavor
\cite{Jaffe:1991ra}.  The operators involving gluons do appear in the
relations between the second moments,
\begin{eqnarray}
  \label{second-moments}
A_{S 20} - A_{G 20} &=& \frac{m_q}{m}\, A_{20}
  - \frac{t}{4m^2}\, \Big( B_{T 20} + 2 \tilde{A}_{T 20} \Big) ,
\nonumber \\
B_{S 20} - B_{G 20} &=& \frac{m_q}{m}\, B_{20}
  - \Big( A_{T 20} - \frac{t}{2m^2} \tilde{A}_{T 20} \Big) ,
\qquad\qquad
C_{S 2} - C_{G 2} \;=\;  \frac{m_q}{m}\, C_{2} \, ,
\nonumber \\
\tilde{A}_{P 21} - \tilde{A}_{G 21} &=&
  \frac{t}{4m^2}\, \tilde{B}_{T 21} \, ,
\nonumber \\
\tilde{B}_{P 20} - \tilde{B}_{G 20} &=& 2 A_{T 20} \, ,
\qquad\qquad
\tilde{B}_{P 22} - \tilde{B}_{G 22} \;=\; - \tilde{B}_{T 21} \, .
\phantom{\frac{t}{2m^2}}
\end{eqnarray}

In forward limit $t=0$ no connection is obtained between moments of
the twist-three distribution $e(x)$ and moments of the transversity
distribution $h_1(x)$.  Rather, form factors of the twist-three
operators which survive in the forward limit are connected with form
factors of the quark tensor current which decouple in that limit, and
vice versa.  Preliminary results on $A_{T n0}$, $\tilde{A}_{T n0}$ and
$B_{T n0}$ from lattice calculations \cite{QCDSF:2005nn} suggest that
these form factors are rather large.  If confirmed, this would imply
that the twist-three combination $A_{S n0} - A_{G n0}$ is quark mass
suppressed at $t=0$ but no longer small for $-t \sim m^2$, and that
the form factor combinations $B_{S n0} - B_{G n0}$ and $\tilde{B}_{P
n0} - \tilde{B}_{G n0}$ are already large at $t=0$.  In other words,
away from $t=0$ chiral-odd twist-three matrix elements would not
generically be small compared with twist-two matrix elements.

The relations (\ref{first-moments}) and (\ref{second-moments}) may be
of practical use in lattice QCD calculations.  Note that the form
factors on the left-hand sides belong to operators with one covariant
derivative more than those on the right-hand sides (counting the gluon
field strength as the commutator of two covariant derivatives).
Operators with more derivatives are less localized on the lattice and
thus more affected by errors.  The form factors on the right-hand
sides of (\ref{first-moments}) and (\ref{second-moments}) have been or
are being calculated in lattice QCD.  Together with lattice
determinations of the renormalized quark masses (see e.g.\
\cite{Rakow:2004vj} and references therein) one may thus use the
equation of motion constraints to determine the twist-three form
factors in (\ref{first-moments}) and the twist-three form factor
combinations in (\ref{second-moments}).  Alternatively, one may
evaluate the twist-three matrix elements on the lattice and use
(\ref{first-moments}) and (\ref{second-moments}) as constraints to
reduce the errors in the extracted form factors.  Note that a separate
determination of $A_{G ni}$, $B_{G ni}$, $\tilde{A}_{G ni}$,
$\tilde{B}_{G ni}$ would allow one to check the often-used
Wandzura-Wilczek approximation, which assumes that matrix elements of
(chiral-even or chiral-odd) quark-antiquark-gluon operators are small.


\section{Summary}
\label{sec:sum}

Generalized transversity distributions at zero skewness $\xi$ describe
the density of transversely polarized quarks in the impact parameter
plane.  We have derived the corresponding expression
(\ref{trans-distr}) and analyzed its detailed structure.  The momentum
space distributions $H_T$, $E_T$ and $\tilde{H}_T$ at $t=0$ and
$\xi=0$ describe simple average features of this density according to
(\ref{b-int-1b}), (\ref{b-int-3}), (\ref{b-int-4}).  The formulae for
the impact parameter density of quarks closely resemble those for
transverse momentum dependent distributions.  This resemblance
exhibits for instance a correspondence between the Sivers function
$f_{1T}^\perp$ and the nucleon helicity-flip distribution $E$, and
between the Boer-Mulders function $h_1^\perp$ and $E_T+2\tilde{H}_T$.
It would be interesting to investigate the correspondence between
impact parameter and transverse momentum dependent distributions at a
dynamical level, as has been done for $f_{1T}^\perp$ and $E$ in
\cite{Burkardt:2003uw,Burkardt:2003je}.  The distribution of quarks in
the impact parameter plane is no longer rotationally symmetric as soon
as either the proton or the quark are transversely polarized, and the
preferred direction of transverse quark polarization is not isotropic
even in an unpolarized proton.  Preliminary results of lattice QCD
calculations suggest that such effects may be quite large.

The impact parameter density of quarks for arbitrary polarization can
be obtained from the spin matrix (\ref{spinm0}).  This matrix is
positive semidefinite, which leads to simple bounds on generalized
parton distributions, as special cases of the general results in
\cite{Pobylitsa:2002iu}.  The most stringent inequalities hold in
impact parameter space.  A combination of $H_T$ and $\tilde{H}_T$ is
bounded by $H$ according to~(\ref{bound2}), and (\ref{bound2a})
extends the bound (\ref{bound1}) previously given by Burkardt
\cite{Burkardt:2003ck}.  The size of the chiral-odd distributions thus
has consequences also in the purely chiral-even sector, since it
restricts the possibilities to saturate the inequality (\ref{bound1}),
which involves only $E$, $H$ and $\tilde{H}$.  By suitable integration
over the impact parameter, one obtains bounds in momentum space.
Bounds can also be given for Mellin moments that correspond to the
\emph{sum} of quark and antiquark distributions.  Since the axial
current has different charge conjugation parity than the vector and
tensor currents, this requires bounds without the quark helicity
distribution~$\tilde{H}$, like (\ref{bound5}), (\ref{bound5a}) and
(\ref{bound2b}), (\ref{bound3b}).  Such bounds can for instance be
applied to the results of lattice QCD calculations.  It will be
interesting to see by how much bounds are violated for Mellin moments
corresponding to the \emph{difference} of quark and antiquark
distributions, since this is a measure for the importance of antiquark
contributions in these moments.

The divergence and the curl of the vector field $F_T^i(x,\tvec{b})$,
which describes transverse quark polarization in the impact parameter
plane, are matrix elements of the total derivatives of twist-two
quark-antiquark operators.  These derivative operators are related to
twist-three operators via the QCD equations of motion, namely to
scalar or pseudoscalar quark currents and to quark-antiquark-gluon
operators.  Such relations have been investigated for forward parton
distributions and for meson distribution amplitudes in the literature.
In (\ref{first-moments}) and (\ref{second-moments}) we give the
corresponding relations for the form factors parameterizing the first
two Mellin moments of generalized parton distributions.  This can
easily be extended to higher moments.  Such relations may be of use
for exploring the twist-three sector in lattice QCD calculations.


\section*{Acknowledgments} 

We gratefully thank A.~Bacchetta, M.~G\"ockeler, P.~Mulders,
A.~Sch{\"a}fer, H.~Wittig and J.~Zanotti for discussions, and
D.~Br{\"o}mmel, P.~Mulders and A.~Sch{\"a}fer for useful remarks on
the manuscript.  This work is supported by the Helmholtz Association,
contract number VH-NG-004, and the 
Integrated Infrastructure Initiative "Study of
strongly interacting matter" of the European Union 
under contract number RII3-CT-2004-506078.



\begin{thebibliography}{99}

\bibitem{Vogelsang:2003eb}
W.~Vogelsang,
hep-ph/0309295;
%
A.~Metz,
hep-ph/0412156;
%
V.~Barone,
hep-ph/0502108.

\bibitem{Goeke:2001tz}
K.~Goeke, M.~V.~Polyakov and M.~Vanderhaeghen,
Prog.\ Part.\ Nucl.\ Phys.\  {\bf 47}, 401 (2001)
[hep-ph/0106012].

\bibitem{Diehl:2003ny}
M.~Diehl,
Phys.\ Rept.\  {\bf 388}, 41 (2003)
[hep-ph/0307382].

\bibitem{Belitsky:2005qn}
A.~V.~Belitsky and A.~V.~Radyushkin,
hep-ph/0504030.

\bibitem{Burkardt:2000za}
M.~Burkardt,
Phys.\ Rev.\ D {\bf 62}, 071503 (2000),
Erratum-ibid.\ D {\bf 66}, 119903 (2002)
[hep-ph/0005108].

\bibitem{Burkardt:2002hr}
M.~Burkardt,
Int.\ J.\ Mod.\ Phys.\ A {\bf 18}, 173 (2003)
[hep-ph/0207047].

\bibitem{Collins:1997fb}
J.~C. Collins, L.~Frankfurt, and M.~Strikman,
Phys.\ Rev.\ D {\bf 56}, 2982 (1997)
[hep-ph/9611433].

\bibitem{Hoodbhoy:1998vm}
P.~Hoodbhoy and X.~D.~Ji,
Phys.\ Rev.\ D {\bf 58}, 054006 (1998)
[hep-ph/9801369].

\bibitem{Diehl:2001pm}
M.~Diehl,
Eur.\ Phys.\ J.\ C {\bf 19}, 485 (2001)
[hep-ph/0101335].

\bibitem{Ivanov:2002jj}
D.~Y.~Ivanov, B.~Pire, L.~Szymanowski and O.~V.~Teryaev,
Phys.\ Lett.\ B {\bf 550}, 65 (2002)
[hep-ph/0209300].

\bibitem{Gockeler:2003jf}
M.~G\"ockeler {\it et al.} [QCDSF Collaboration],
Phys.\ Rev.\ Lett.\  {\bf 92}, 042002 (2004)
[hep-ph/0304249].

\bibitem{Hagler:2003jd}
Ph.~H\"agler {\it et al.} [LHPC Collaboration],
Phys.\ Rev.\ D {\bf 68}, 034505 (2003)
[hep-lat/0304018];\\
%
P.~H\"agler {\it et al.} [LHPC Collaboration],
Phys.\ Rev.\ Lett.\  {\bf 93}, 112001 (2004)
[hep-lat/0312014].

\bibitem{Gockeler:2005aw}
M.~G\"ockeler {\it et al.} [QCDSF Collaboration],
hep-lat/0501029.

\bibitem{Ji:1998pc}
X.~D.~Ji,
J.\ Phys.\ G {\bf 24}, 1181 (1998)
[hep-ph/9807358].

\bibitem{Soper:1976jc}
D.~E.~Soper,
Phys.\ Rev.\ D {\bf 15}, 1141 (1977).

\bibitem{Diehl:2002he}
M.~Diehl,
Eur.\ Phys.\ J.\ C {\bf 25}, 223 (2002),
Erratum-ibid.\ C {\bf 31}, 277 (2003)
[hep-ph/0205208].

\bibitem{Soper:1972xc}
D.~E.~Soper,
Phys.\ Rev.\ D {\bf 5}, 1956 (1972).

\bibitem{Boer:2003cm}
D.~Boer, P.~J.~Mulders and F.~Pijlman,
Nucl.\ Phys.\ B {\bf 667}, 201 (2003)
[hep-ph/0303034].

\bibitem{Boglione:1999pz}
M.~Boglione and P.~J.~Mulders,
Phys.\ Rev.\ D {\bf 60}, 054007 (1999)
[hep-ph/9903354].

\bibitem{Mulders:1995dh}
P.~J.~Mulders and R.~D.~Tangerman,
Nucl.\ Phys.\ B {\bf 461}, 197 (1996),
Erratum-ibid.\ B {\bf 484}, 538 (1997)
[hep-ph/9510301].

\bibitem{Boer:1997nt}
D.~Boer and P.~J.~Mulders,
Phys.\ Rev.\ D {\bf 57}, 5780 (1998)
[hep-ph/9711485].

\bibitem{Collins:2002kn}
J.~C.~Collins,
Phys.\ Lett.\ B {\bf 536}, 43 (2002)
[hep-ph/0204004].

\bibitem{Burkardt:2003uw}
M.~Burkardt,
Nucl.\ Phys.\ A {\bf 735}, 185 (2004)
[hep-ph/0302144].

\bibitem{Burkardt:2003je}
M.~Burkardt and D.~S.~Hwang,
Phys.\ Rev.\ D {\bf 69}, 074032 (2004)
[hep-ph/0309072].

\bibitem{Sachs:1962aa}
R.~G. Sachs,
Phys. Rev. {\bf 126}, 2256 (1962).

\bibitem{Belitsky:2003nz}
A.~V.~Belitsky, X.~D.~Ji and F.~Yuan,
Phys.\ Rev.\ D {\bf 69}, 074014 (2004)
[hep-ph/0307383].

\bibitem{QCDSF:2005nn}
QCDSF Collaboration, in preparation.

\bibitem{Bacchetta:1999kz}
A.~Bacchetta, M.~Boglione, A.~Henneman and P.~J.~Mulders,
Phys.\ Rev.\ Lett.\  {\bf 85}, 712 (2000)
[hep-ph/9912490];\\
%
A.~Bacchetta, M.~Boglione, A.~Henneman and P.~J.~Mulders,
hep-ph/0005140.

\bibitem{Burkardt:2003ck}
M.~Burkardt,
Phys.\ Lett.\ B {\bf 582}, 151 (2004)
[hep-ph/0309116].

\bibitem{Diehl:2004cx}
M.~Diehl, T.~Feldmann, R.~Jakob and P.~Kroll,
Eur.\ Phys.\ J.\ C {\bf 39}, 1 (2005)
[hep-ph/0408173].

\bibitem{Pobylitsa:2002ru}
P.~V.~Pobylitsa,
Phys.\ Rev.\ D {\bf 70}, 034004 (2004)
[hep-ph/0211160].

\bibitem{Pobylitsa:2002iu}
P.~V.~Pobylitsa,
Phys.\ Rev.\ D {\bf 66}, 094002 (2002)
[hep-ph/0204337].
  
\bibitem{Belitsky:2000vx}
A.~V.~Belitsky and D.~M{\"u}ller,
Nucl.\ Phys.\ B {\bf 589}, 611 (2000)
[hep-ph/0007031].

\bibitem{Jaffe:1991ra}
R.~L.~Jaffe and X.~D.~Ji,
Nucl.\ Phys.\ B {\bf 375}, 527 (1992).

\bibitem{Kodaira:1998jn}
J.~Kodaira and K.~Tanaka,
Prog.\ Theor.\ Phys.\  {\bf 101}, 191 (1999)
[hep-ph/9812449].

\bibitem{Ball:1998sk}
P.~Ball, V.~M.~Braun, Y.~Koike and K.~Tanaka,
Nucl.\ Phys.\ B {\bf 529}, 323 (1998)
[hep-ph/9802299].

\bibitem{Ball:1998je}
P.~Ball,
JHEP {\bf 9901}, 010 (1999)
[hep-ph/9812375].

\bibitem{Jaffe:1996zw}
R.~L.~Jaffe,
hep-ph/9602236.

\bibitem{Hagler:2004yt}
Ph.~H{\"a}gler,
Phys.\ Lett.\ B {\bf 594}, 164 (2004)
[hep-ph/0404138].

\bibitem{Efremov:2002qh}
A.~V.~Efremov and P.~Schweitzer,
JHEP {\bf 0308}, 006 (2003)
[hep-ph/0212044].

\bibitem{Rakow:2004vj}
P.~E.~L.~Rakow,
Nucl.\ Phys.\ Proc.\ Suppl.\  {\bf 140}, 34 (2005)
[hep-lat/0411036].

\end{thebibliography}
\end{document}